\documentclass[aps,twocolumn,showpacs,amsmath,amssymb,superscriptaddress]{revtex4}

\usepackage{epsfig}
\usepackage{graphicx}% Include figure files
\usepackage{dcolumn}% Align table columns on decimal point
\usepackage{bm}% bold math
\usepackage{hyperref} % hyperreferences

\begin{document}

\title{Binary black hole initial data from matched asymptotic expansions}

\author{Nicol\'as  Yunes}
\affiliation{Institute for Gravitational Physics and Geometry,
             Center for Gravitational Wave Physics,
             Department of Physics, The Pennsylvania State University,
             University Park, PA 16802-6300}

\author{Wolfgang Tichy}
\affiliation{Institute for Gravitational Physics and Geometry,
             Center for Gravitational Wave Physics,
             Department of Physics, The Pennsylvania State University,
             University Park, PA 16802-6300}
\affiliation{Department of Physics, Florida Atlantic University,
             Boca Raton, FL  33431}

\author{Benjamin J. Owen}
\affiliation{Institute for Gravitational Physics and Geometry,
             Center for Gravitational Wave Physics,
             Department of Physics, The Pennsylvania State University,
             University Park, PA 16802-6300}

\author{Bernd Br\"ugmann}
\affiliation{Institute for Gravitational Physics and Geometry,
             Center for Gravitational Wave Physics,
             Department of Physics, The Pennsylvania State University,
             University Park, PA 16802-6300}
\affiliation{Physikalisch-Astronomische Fakult\"at, 
             Friedrich-Schiller-Universit\"at Jena, 07743 Jena, Germany}

\date{$$Id: paper.tex,v 1.186 2006/09/22 20:36:05 yunes Exp $$}


% Sometimes we want to include preprint numbers, let's put them here
\preprint{IGPG-04/10-6}

%-------------------------------------------------------------------------
%Useful Definitions
%------------------------------------------------------------------------
%
\newcommand\be{\begin{equation}}
\newcommand\ba{\begin{eqnarray}}
\newcommand\ee{\end{equation}}
\newcommand\ea{\end{eqnarray}}
\newcommand\p{{\partial}}
%

%-------------------------------------------------------------------------
\begin{abstract}
%-----------------------------------------------------------------------
  
  We present an approximate metric for a binary black hole spacetime
  to construct initial data for numerical relativity.  This metric is
  obtained by asymptotically matching a post-Newtonian metric for a
  binary system to a perturbed Schwarzschild metric for each hole. In
  the \textit{inner zone} near each hole, the metric is given by the
  Schwarzschild solution plus a quadrupolar perturbation corresponding
  to an external tidal gravitational field.  In the \textit{near
    zone}, well outside each black hole but less than a reduced
  wavelength from the center of mass of the binary, the metric is
  given by a post-Newtonian expansion including the lowest-order
  deviations from flat spacetime.  When the near zone overlaps each
  inner zone in a \textit{buffer zone}, the post-Newtonian and
  perturbed Schwarzschild metrics can be asymptotically matched to
  each other.  By demanding matching (over a $4$-volume in the buffer
  zone) rather than patching (choosing a particular $2$-surface in the
  buffer zone), we guarantee that the errors are small in all zones.
  The resulting piecewise metric is made formally
  $C^\infty$ with smooth transition functions so as to obtain
  the finite extrinsic curvature of a $3$-slice. 
  In addition to the metric and extrinsic curvature, we present
  explicit results for the lapse and the shift,
  which can be used as initial data for numerical simulations.  This
  initial data is not accurate all the way to the asymptotically flat ends
  inside each hole, and therefore must be used with evolution codes which
  employ black hole excision rather than puncture methods.
  This
  paper lays the foundations of a method that can be straightforwardly
  iterated to obtain initial data to higher perturbative order.

%-----------------------------------------------------------------------
\end{abstract}
%-----------------------------------------------------------------------

\maketitle

%%%%%%%%%%%%%%%%%%%%%%%%%%%%%%%%%%%%%%%%%%%%%%%%%%%%%%%%%%%
\section{Introduction}
%%%%%%%%%%%%%%%%%%%%%%%%%%%%%%%%%%%%%%%%%%%%%%%%%%%%%%%%%%%

The simulation of binary black-hole systems is of fundamental physical
interest as the purely general relativistic two-body problem.  It is
also of astrophysical interest, since accurate simulations of the late
inspiral and merger phases of such binaries will considerably help the
effort to detect the gravitational-wave signals and extract
information from them~\cite{Baumgarte:2002jm}.  Simulation reduces to
the numerical solution of the Cauchy problem: take some initial data
and evolve it.  The evolution is difficult for many reasons, although
in recent years there has been much progress. Still, any evolution is
only as good as its initial data.

A key issue of initial data is astrophysical realism. The goal is to
compute data on a hypersurface that represents one moment in time of
an astrophysical inspiral of two black holes. If such an inspiral is
defined by initial conditions in the distant past for widely separated
black holes, then the only way to obtain the exact data at a later
time would be to perform the actual evolution using the full Einstein
equations. This procedure, however, is computationally expensive and
thus impractical. On the other hand, several schemes have been
developed to pose initial data that approximates the astrophysical
situation at a given time. These schemes are typically either based on
post-Newtonian (PN) methods or on the numerical solution of the
constraint equations of relativity.

For example, the literature provides many types of initial data for
black holes in approximately circular
orbits~\cite{Cook94,Matzner98a,Baumgarte00a,
Marronetti00,Cook:2000vr,Grandclement02,Pfeiffer:2002xz,Baker02a,
Tichy:2002ec,Tichy03a,Tichy:2003qi,Yo:2004ng,Cook:2004kt} that satisfy
the constraints of the Einstein equations. To obtain such data, certain
assumptions are made, such as conformal flatness
and quasi-circularity. These assumptions are expected to be good
approximations within a certain error, although the astrophysical
metric after a long inspiral is known to be not exactly conformally
flat and the orbit not perfectly circular.

In this paper we consider a post-Newtonian method combined with black
hole perturbation theory to construct approximate inspiral initial
data. For large to intermediate separations of compact objects, an
astrophysically relevant approximate spacetime can be obtained far
from the black holes by analytical post-Newtonian and post-Minkowskian
methods~\cite{Blanchet:2002av}. One advantage of such methods is that
they allow systematic improvements through higher order
expansions (compared to numerical constraint solving schemes which typically
include only the correct lowest order PN behavior). In their regime
of validity, PN methods do result in appropriate deviations from
conformal flatness and in non-circular inspiral orbits. 
The main disadvantage of PN methods is that, by construction, they are
generally believed to fail in the final phase of the inspiral for fast
moving objects, and also close to non-pointlike objects with horizons.
On the other hand, black hole perturbation theory provides an accurate
spacetime in a region of the manifold sufficiently close to the background
black hole. The main disadvantage of this theory is that it fails
sufficiently far from the background hole and, thus, cannot provide
information about the dynamics of the entire spacetime.

In what follows, we take a concrete step toward combining PN theory with
black hole perturbation theory using the mathematical machinery of
asymptotic matching.
The method maintains the potential for systematic improvement through
higher order expansions, although we only work at low order here~\cite{Yunes:2006iw,higher-order-matching}.
From the PN approach the method inherits its astrophysical justification,
\textit{i.e.}\ that for sufficient separation between the holes the method
will yield metric components that are correct up to uncontrolled remainders
of certain orders.
The uncontrolled remainders in the metric components have different effects
on different quantities, and it is of interest to see how other quantities
such as the binding energy compare to those of other initial data sets in
the literature.
This question is beyond the scope of this article, but should be
addressed in the future.

Concretely, we have to discuss how black holes are incorporated in our
approach. While formally PN methods assume slow motion and weak
internal gravity of the sources, it has been shown that the results
hold as well for objects such as black holes with strong internal
gravity~\cite{Thorne:1984mz} as long as one is not too close to these objects.
Near each black hole, a tidally
perturbed Schwarzschild or Kerr spacetime provides another analytical
approximation.  Given that different approximate metrics can be
constructed from different scale expansions, it is natural to try the
method of matched asymptotic expansions~\cite{Bender}.  If there is an
overlap region (also known as a buffer zone)
where both approximations (post-Newtonian and tidal
perturbation) are valid, a diffeomorphism can be constructed between
charts used in different regions of the manifold by different
approximation schemes.  Matching---demanding that both approximation
schemes have the same asymptotic form in the overlap region---relates
physical observables in the different regions, \textit{i.e.}\ ensures
that both expansions represent the same physical system.

The first attempt to construct initial data in such a way was by
Alvi~\cite{Alvi:1999cw}.
By construction, there are discontinuities in the data, which were found to
be too large for numerical experiments~\cite{Bernd}.
Alvi's fundamental problem was that, in the terminology of textbooks such
as Bender's~\cite{Bender}, he did not \textit{match} (construct expansions
asymptotic to each other everywhere in the overlap region) but rather
\textit{patched} (set approximate solutions to the Einstein equations
equal to each other on specified
2-surfaces) so that large errors in the extrinsic curvature were possible.
Alvi's Table~I shows that his spatial metric near the black holes is
discontinuous apart from the Minkowski terms (independent of $G$) in
either region.
Such discontinuities are problematic for numerical relativity, since part
of the initial data is the extrinsic curvature which includes derivatives
of the spacetime metric.
Smoothing can be attempted, for example with transition functions as in
Alvi's next paper~\cite{Alvi:2003pn}, but it is not trivial to
implement---especially with such large discontinuities---without adding
unphysical content to the initial data.
There is also the issue of making sure that the initial data slicing is
treated consistently in the various expansions, which Alvi addressed to
some extent but did not always make explicit his assumptions.
Finally, there was a problem with the accuracy to which Alvi
calculated metric components. Construction of the extrinsic curvature
requires terms in the expansions that Alvi did not calculate because he
assumed (incorrectly) that
the counting of orders follows the standard pattern used in deriving
post-Newtonian equations of motion.

The main point of this paper is that we are able to correct the
mathematical problems with Alvi's approach, and that we provide
initial data for actual numerical evolutions. We use true asymptotic
matching to construct a piecewise metric for two black holes in
circular orbit, including terms of order the gravitational constant
$G$ on the diagonal of the metric and $O(G)^{3/2}$
off the diagonal. We then remove the piecewise nature of the
approximate metric by ``merging'' or ``smoothing'' the solutions in
the buffer zones, thus generating a uniform approximate metric. We do
this by constructing smooth transition functions so that the uniform
approximate solution is in principle $C^{\infty}$, although in
practice higher-order derivatives will be less accurate than
lower-order ones. These transition functions are carefully constructed
to avoid introducing errors in the smoothed global metric larger than
those already contained in the approximate solutions. This metric
allows for the calculation of the lapse to $O(G)$, the shift to
$O(G)^{3/2}$, and the extrinsic curvature to $O(G)^{3/2}$. Although
this data contains only the first order deviations from flat
spacetime, our approach does include the tidal perturbations near the
black holes.
Strictly speaking, these tidal perturbations are valid only near the
horizons---our approach cannot model 
the asymptotically flat ends inside the holes and therefore must be used in
numerical evolutions with excision rather than punctures.
Our formalism can be extended to higher order by including more precise
post-Newtonian~\cite{Blanchet:2004ek} and black hole perturbation
theory results~\cite{Poisson:2005pi} already in the literature.

By construction, our initial data satisfies the
constraints only up to uncontrolled remainders of a certain order.
Therefore, this data may still need to be projected to the constraint
hypersurface via a conformal decomposition.
One avenue worth exploring is that since the constraint
violations are $O(G)^{3/2}$ or smaller, it may be possible to find a
constraint projection algorithm that changes the physical content of our
initial data only at a comparably small order. In this manner, the
formalism presented here can potentially be used to construct
extremely accurate background data for constraint solving. This hybrid
combination of an accurate background $4$-metric plus constraint
solving might potentially lead to very astrophysically realistic
initial data, which then could be compared and tested against other sets
already in the literature.

This paper is organized as follows: Sec.~\ref{scenario} describes the
method of asymptotic matching as applicable to this problem.
Sec.~\ref{near} discusses the near zone expansion of the metric and
determines its asymptotic expansion in the overlap region.
Sec.~\ref{inner} concentrates on the inner zone expansion of the metric
and expands
it asymptotically in the overlap region.  Sec.~\ref{matching} applies
asymptotic matching to the metrics to obtain matching relations
between expansion coefficients and a map that relates the charts used
in the different regions.  Sec.~\ref{secglobal} constructs the global
metric, discusses its properties, and builds transition functions to
eliminate discontinuities between local approximations.
Sec.~\ref{numerics} computes the extrinsic curvature, lapse and shift.
Sec.~\ref{conclusion} concludes and points toward future research.

Throughout this paper we use geometrized units ($G=c=1$) and we have
relied heavily on symbolic manipulation software, such as MAPLE and
MATHEMATICA. We use the tilde as a relational symbol such that $a \sim
b$ means ``$a$ is asymptotic to $b$''~\cite{Bender}.
When we refer to our results as ``global,'' we mean that they cover the
region of most interest to numerical relativity.
Strictly speaking, our results do not cover the radiation zone (further
from the binary than a reduced wavelength) or the asymptotically flat ends
inside the holes.
However, obtaining the radiation-zone solution and matching it to the
near-zone solution is a solved problem~\cite{Blanchet:2002av} and the
asymptotic ends inside the holes can be removed with excision before
numerically evolving the initial data.

%%%%%%%%%%%%%%%%%%%%%%%%%%%%%%%%%%%%%%%%%%%%%%%%%%%%%%%%%
\section{Approximation Regions and Precision}
\label{scenario}
\begin{figure}
\includegraphics[scale=0.33]{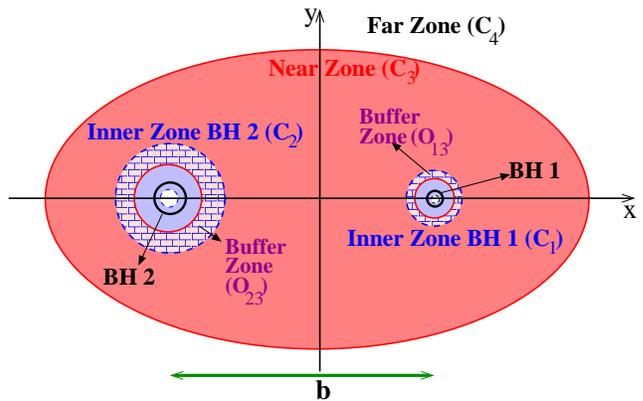}
\caption{\label{egg}
Schematic diagram of the near zone (dark gray), inner zones (light
gray) and buffer zones (checkered) projected onto the orbital 
plane.
The black holes' horizons are shown by solid black lines, while the
excision boundaries are shown by dashed black lines.  
The near zone overlaps the inner zone of each black hole, and these overlap
regions are the buffer zones (checkered patterns.)
The boundaries of all zones are somewhat imprecise since they are based on
power series approximations, but the buffer zones are roughly spherical
shells shown in this Figure as annuli.
}
\end{figure}

%%%%%%%%%%%%%%%%%%%%%%%%%%%%%%%%%%%%%%%%%%%%%%%%%%%%%%%%%
Let us now consider a binary black hole spacetime, with holes of mass
$m_1$ and $m_2$, total mass $m = m_1 + m_2$ and spatial coordinate separation
$b$. The manifold (Fig.~\ref{egg}) can be divided into $4$
submanifolds, the boundaries of which cannot be determined precisely
due to the presence of uncontrolled remainders in black hole
perturbation theory (BHPT) and post-Newtonian (PN) asymptotic series.
Nonetheless, an approximate subdivision is possible and we make one as
follows:
\begin{enumerate}
\item {\bf{The inner zone of Black Hole $1$}}, (submanifold
  ${\cal{C}}_1$): $R_1 \ll b $, where $R_A$ is the distance from the
  $A$th black hole in isotropic coordinates. In this region, the
  metric is obtained via black hole 
  perturbation theory as an expansion in $\epsilon_{(1)} = R_1/b$.
  \cite{Alvi:1999cw,Thorne:1984mz}.
\item {\bf{The inner zone of Black hole $2$}}, (submanifold
  ${\cal{C}}_2$): $R_2 \ll b$, where the metric is obtained in the
  same manner as in region ${\cal{C}}_1$ but with labels 1 and 2
  swapped.
\item {\bf{The near zone}}, (submanifold ${\cal{C}}_3$): $r_A \gg m_A
  $ and $r \leq \lambda/2 \pi$, where $\lambda$ is the wavelength of
  gravitational radiation, $r$ is the distance from the binary center
  of mass in harmonic coordinates, and $r_A-m_A$ is the separation in
  harmonic coordinates from the horizon of the $A$th black hole.  In
  this region, a post-Newtonian approximation is used for the metric
  with an expansion parameter $\epsilon_{(3)} = m_A/r_A$
  \cite{Blanchet:2002av} which is formally treated as the same order
  for both values of $A$.
\item
{\bf{The far zone}}, (submanifold ${\cal{C}}_4$): $r \geq \lambda/2
\pi $, where the metric is obtained from a post-Minkowski calculation
\cite{Will:1996zj}.
\end{enumerate}
These zones are shown schematically in Fig.~\ref{egg}, projected onto
the orbital plane. In
these figures, the near zone is shown in dark gray, the inner zones in
light gray
and the buffer zones in a checkered pattern. The holes' horizons are denoted by 
black solid lines, while the dashed black lines are the excision
boundaries. 
%

% The dependence of the approximations...
The manifold is subdivided in such a way so that approximate solutions
to the Einstein equations can be found in each region. These
approximate solutions will depend on certain coordinates and
parameters local only to that region. The near zone metric, $g_{\mu
  \nu}^{(3)}$, is an expansion in $\epsilon_{(3)} \equiv m/r \ll 1$,
which depends on harmonic coordinates $x^{\mu}$ and parameters,
such as the masses $m_A$ of the holes and
the angular velocity $\omega$ of the system. Similarly, the metric in
inner zone $1$ (or $2$), $h_{\bar{\mu}\bar{\nu}}^{(1)}$ (or
$h_{\bar{\mu}\bar{\nu}}^{(2)}$), is an expansion in $\epsilon_{(1)}
\equiv r_{1}/b \ll 1$ (or $\epsilon_{(2)} \equiv r_{2}/b \ll 1$),
which depends on isotropic coordinates $x^{\bar{\mu}}$ and certain
parameters, such as the mass of the background hole $M$
and the angular velocity $\Omega$ of the tidal perturbation. The parameters and
the coordinates used in different regions are not identical and
are valid only inside their respective regions, although those regions
overlap. 

% A little history on matching...
A global metric can be obtained by relating the different approximate
solutions through asymptotic matching. The theory of matched asymptotic
expansions was first developed to perform multiple scale analysis on
non-linear partial differential equations and to obtain global
approximate solutions~\cite{Bender}.  In general relativity, this
method was first applied by Burke and Thorne~\cite{Burke-Thorne},
Burke~\cite{Burke}, and D'Eath~\cite{Death:1975,Death:1974o} in the
$1970$s to derive corrections to the laws of motion due to coupling of
the body's motion to the geometry of the surrounding spacetime. Based
on these ideas, in this paper we will develop a version of the theory
of matched asymptotics that is useful to obtain initial data for
numerical relativity simulations.

% The buffer zones...
Asymptotic matching consists of relating different approximate
solutions inside a common region of validity.
This region is usually called the buffer zone by relativists, but is called
the overlap region by others.
For a binary there exists three such regions, which are $4$-volumes:
Two buffer zones (${\cal{O}}_{13}$ and ${\cal{O}}_{23}$) are
defined by the intersection of the near zone and the inner zones of
black hole $1$ and $2$; the third one is defined by the intersection
of the near zone and the radiation zone (${\cal{O}}_{34}$). The former
two, shown in Fig.~\ref{egg} in a checkered pattern, are defined
by the asymptotic condition $m_A \ll r_A \ll b$. The latter
has been analyzed in Ref.~\cite{Blanchet:2002av} and will not
be discussed here. In this paper we perform asymptotic matching in the
former two buffer zones $\mathcal{O}_{13}$ and $\mathcal{O}_{23}$.
In order for our tidal perturbations in the inner zones to be valid, the
inner zones $\mathcal{C}_1$ and $\mathcal{C}_2$ cannot overlap.

% The matching condition...
Once a buffer zone has been found, asymptotic matching can be used to
relate adjacent approximate solutions. The first step is to find the
asymptotic expansions of the approximate solutions inside the buffer
zones. These approximate solutions depend on the expansion
parameters, $\epsilon_{(1)}$, $\epsilon_{(2)}$ and $\epsilon_{(3)}$,
which are small only in their respective regions of validity
$\mathcal{C}_1$, $\mathcal{C}_2$, and $\mathcal{C}_3$.
By definition, in each overlap region both expansion parameters are small,
specifically ${\epsilon_{(1)}} \ll 1$ and
$\epsilon_{(3)} \ll 1 $ in ${\cal{O}}_{13}$, while ${\epsilon_{(2)}}
\ll 1$ and $\epsilon_{(3)} \ll 1$ in ${\cal{O}}_{23}$.  Inside buffer
zone $1$, for example, we can then asymptotically expand the near zone
solution in $\epsilon_{(1)} \ll 1$ to obtain $\tilde{g}_{\mu
  \nu}^{(3)}$ and the inner zone solution in ${\epsilon_{(3)}} \ll 1$
to obtain $\tilde{h}_{\bar{\mu} \bar{\nu}}^{(1)}$. These asymptotic
expansions of approximate solutions are naturally bivariate since they
depend on two {\emph{independent}} expansion parameters.
When working
with these bivariate expansions, we use the symbol $O(p,q)$ both
to denote terms of order $(m/b)^p (r_A/b)^q$ and uncontrolled
remainders of order $(m/b)^p$ or $(r_A/b)^q$. Relating adjacent
approximations then reduces to imposing the asymptotic matching
condition
\be
\tilde{g}_{\mu \nu}^{(3)} \sim \tilde{h}_{\bar{\mu}\bar{\nu}}
\frac{\partial x^{\bar{\mu}}}{\partial x^{\mu}} \frac{\partial
  x^{\bar{\nu}}}{\partial x^{\nu}}. 
\label{matchingcondition}
\ee
This expression means not that the two approximate solutions are equated,
which is correctly called ``patching,'' but rather that all coefficients of
all controlled terms in the bivariate expansions are equated.

% The piecewise metric...
After imposing the asymptotic matching condition, one obtains a
coordinate and parameter transformation between the near zone and the
inner zone $1$ in ${\cal{O}}_{13}$ (and similarly in
${\cal{O}}_{23}$.) These transformations allow for the construction of
a global piecewise metric, which is guaranteed to be asymptotically
smooth in the buffer zone up to uncontrolled remainders in the
matching scheme. Asymptotic smoothness here means that adjacent pieces
of the piecewise global metric and all of their derivatives are
asymptotic to each other inside the buffer zones. This asymptotic
smoothness, however, does not rule out small discontinuities on
the order of the uncontrolled remainders in the approximations,
when we pass from one approximation to the other. The global metric
can be made formally $C^{\infty}$ by smoothing over these discontinuities,
which introduces a new error into the solution. Asymptotic smoothness,
however, guarantees that this error will be smaller than or equal to that
already contained in the uncontrolled remainders of the
approximations, provided the smoothing functions are sufficiently
well-behaved.

%Order enumeration...
Finally, we enumerate the orders of approximation used in the near
zone.  The Einstein equations are guaranteed to generate a well-posed
initial value problem for globally hyperbolic
spacetimes~\cite{Baumgarte:2002jm}, where the initial data could
consist of the extrinsic curvature $K_{ij}$ and the spatial $3$-metric
$\gamma_{ij}=g_{ij}$.  We can write the extrinsic curvature in the
form
\be
K_{ij} = \frac{1}{2 \alpha} \left(2 D_{(i} \beta_{j)} - \partial_t
\gamma_{ij}\right),
\ee
where $\beta_i=g_{0i}$ is the shift vector and $\alpha$ is the lapse.
Time derivatives are smaller than spatial derivatives by a
characteristic velocity, which by the virial theorem is
$O(m/b)^{1/2}$.  Therefore to compute $K_{ij}$ consistently to a given
order in $m/b$, the 4-metric components $g_{0i}$ are needed to
$O(m/b)^{1/2}$ beyond the highest order in $g_{ij}$ (and $g_{00}$,
which appears in $\alpha$).  In this paper we compute the first two nonzero
contributions to the $3$-metric and extrinsic curvature, \textit{i.e.}\ the
leading order terms and the lowest-order corrections.
This means that we need the 4-metric components $g_{00}$ and $g_{ij}$
to $O(m/b)$, but we need $g_{0i}$ to $O(m/b)^{3/2}$.  Note that this
does not correspond to any standard post-Newtonian order counting or
nomenclature, which is why we quote precisions and remainders
precisely in terms of expansion parameters rather than in ambiguous
terms such as ``$n$th PN''. The standard post-Newtonian order counting
corresponds to the calculation of the equations of motion for spinless
bodies, but the counting must be altered when studying other problems,
such as the bending of light or the equations of motion for bodies
with spin~\cite{Tagoshi:2000zg}.

%%%%%%%%%%%%%%%%%%%%%%%%%%%%%%%%%%%%%%%%%%%%%%%%%%%
\section{Near Zone Metric}
\label{near}
%%%%%%%%%%%%%%%%%%%%%%%%%%%%%%%%%%%%%%%%%%%%%%%%%%%

In this section, we present the post-Newtonian (PN) metric in the near
zone ${\cal C}_3$ and expand it in the overlap region $\mathcal{O}_{13}$,
the buffer zone where the inner zone ${\cal C}_1$ of BH1 and the near zone
overlap.
When performing the matching in Sec.~\ref{matching}, we will obtain the
corresponding expansion in the other overlap region $\mathcal{O}_{23}$ by a
simple symmetry transformation.

\begin{figure}[t]
\includegraphics[scale=0.5]{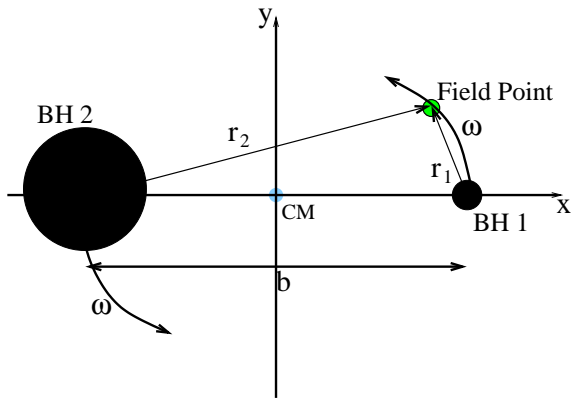}
\caption{\label{bh}
Diagram of the near zone coordinate system used for the post-Newtonian
metric in harmonic coordinates.
The $z$ axis is chosen parallel to the orbital angular momentum of the
binary, so that the black holes orbit counter-clockwise here with angular
velocity $\omega$.
The origin is chosen to be the center of mass, while $r_1$ and $r_2$ denote
the coordinate separations of the field point from hole 1 and 2
respectively.
The coordinate separation between the holes is $b$.
}
\end{figure}

We use harmonic corotating coordinates $(t,x,y,z)$ rotating around the
center of mass such that
\ba
\label{corotatingtransformation}
t'&=&t,
\nonumber \\
x'&=&x\cos{\omega t} - y \sin{\omega t},
\nonumber \\
y'&=&x \sin{\omega t} + y \cos{\omega t},
\nonumber \\
z'&=&z,
\ea
where primed coordinates are nonrotating.
The near-zone metric takes the form~\cite{Blanchet:1998vx}
\ba
g_{00}^{(3)} &\approx& -1 + \frac{2 m_1}{r_1} + \frac{2 m_2}{r_2} 
+ \omega^2 (x^2 + y^2),
\nonumber \\
g_{01}^{(3)} &\approx& -y \; \omega \left( 1 + \frac{2 m_1}{r_1}\right),
\nonumber \\
g_{02}^{(3)} &\approx& x \; \omega \left( 1 + \frac{2 m_1}{r_1}\right) 
- 4 \mu b \omega \left(\frac{1}{r_1} - \frac{1}{r_2}\right), 
\nonumber \\
\label{nearmetricC:1}
g_{ij}^{(3)} &\approx& \delta_{ij} \left(1 + \frac{2m_1}{r_1} +
\frac{2m_2}{r_2}\right),
\ea
where all remainders are at least $O(m/r)^2$ and the superscript reminds us
we are working on submanifold ${\cal{C}}_3$.
Here
\ba
r_1 = \sqrt{x_1^2 + y^2 + z^2} = \sqrt{(x-m_2b/m)^2 + y^2 + z^2} , 
\nonumber \\
r_2 = \sqrt{x_2^2 + y^2 + z^2} = \sqrt{(x+m_1b/m)^2 + y^2 + z^2}
\ea
are the usual harmonic radial coordinates centered on black holes 1 and 2
and $b$ is the separation between holes.
Implicit in this metric is the assumption that $r_A$ (for $A=1,2$) is of
order $b$, or in other words that the field point is not too close to one
of the holes.
(Recall that, in harmonic coordinates, the horizons are at $r_A=m_A$ if we
neglect tidal deformations.)
We do not include the $O(m/r)^2$ terms in $g_{00}$ in what is commonly
called the first post-Newtonian (1PN) metric and we do include
$O(m/r)^{3/2}$ terms in $g_{0i}$ for reasons discussed at the end of
Sec.~\ref{scenario}.
In Eq.~(\ref{nearmetricC:1}) it is sufficient to use the first
post-Newtonian approximation
\be
\omega=\sqrt{\frac{m}{b^3}}
\left[1 - \frac{1}{2} \left(3-\frac{\mu}{m}\right)
  \frac{m}{b} + O\left(\frac{m}{b}\right)^2\right], 
\ee 
to the angular velocity, where $\mu=m_1m_2/(m_1+m_2)$ is the reduced mass
of the binary.
Our choice of sign corresponds to coordinates in which the orbital angular
momentum is in the positive $z$ direction, as shown in Fig.~\ref{bh}.

We now concentrate on the overlap region (buffer zone) ${\cal
O}_{13}$.  Inside ${\cal O}_{13}$ we expand $1/r_2$ as a power series in
$r_1$ as
\be
\label{1/r_2}
\frac{1}{r_2} = \frac{1}{b} \sum_{n=0}^{\infty} (-1)^n \left(\frac{r_1}{b}\right)^{n} 
P_n\left(\frac{x_1}{r_1}\right),
\ee
where the $P_n$ are Legendre polynomials.
Substituting into Eq.~(\ref{nearmetricC:1}), we obtain
\begin{widetext}
\ba
\tilde{g}_{00}^{(3)} &\sim& -1 + \frac{2m_1}{r_1} + 
\frac{2m_2}{b} \left[ 1 - \frac{r_1}{b} P_1\left(\frac{x_1}{r_1}\right) +
\left(\frac{r_1}{b}\right)^2 P_2\left(\frac{x_1}{r_1}\right) \right] + \omega^2 \left(x^2
+ y^2\right),
\nonumber \\
%\\
\tilde{g}_{01}^{(3)} &\sim& - y \; \omega \left\{ 1 + \frac{2 m_1}{r_1}
+ \frac{2 m_2}{b}
\left[1-\frac{r_1}{b}P_1\left(\frac{x_1}{r_1}\right)\right]\right\},
\nonumber \\
%\\
\tilde{g}_{02}^{(3)} &\sim& x \; \omega \left\{ 1 + \frac{2 m_1}{r_1} + \frac{2 m_2}{b}
\left[1-{r_1\over{b}}P_1\left(\frac{x_1}{r_1}\right)\right]\right\}
- 4 \mu b \omega \left\{{1\over{r_1}} - {1\over{b}}\left[1-{r_1\over{b}}P_1\left(\frac{x_1}{r_1}\right) 
+ \left(r_1\over{b}\right)^2 P_2\left(\frac{x_1}{r_1}\right) \right]\right\}, 
\nonumber \\
%\\
\tilde{g}_{03}^{(3)} &\sim& O(2,3),
\nonumber \\
%\\
\label{nearasympt}
\tilde{g}_{ij}^{(3)} &\sim& \delta_{ij} 
\left\{1 + {2m_1\over{r_1}} + {2m_2\over{b}} \left[ 1 - {r_1\over{b}} P_1\left(\frac{x_1}{r_1}\right) +
\left(r_1\over{b}\right)^2 P_2\left(\frac{x_1}{r_1}\right) \right] \right\}, 
\ea
\end{widetext}
where all errors are of order $O(2,3)$ and where $m_1 \ll r_1 \ll b$.
The metric~(\ref{nearasympt}), denoted with a tilde, is the asymptotic
expansion in the inner zone of BH1 of the PN metric, which is already
an asymptotic expansion in the near zone.

Observe that these expansions constitute a series within a series
(bivariate series).  In order to see this more clearly, we can
rearrange the spatial metric to get
\ba
\label{frob}
\tilde{g}_{ij}^{(3)} &\sim& \delta_{ij} 
\left(1 + {2m_1\over{r_1}}  
\left\{ 1 + {m_2\over{m_1}} {r_1\over{b}} \left[ 1 - {r_1\over{b}}
\right. \right. \right.
\\ \nonumber 
&& \left. \left. \left.  P_1\left(\frac{x_1}{r_1}\right) + \left({r_1\over{b}}\right)^2
P_2\left(\frac{x_1}{r_1}\right) \right] \right\} \right),
\\ \nonumber && \qquad  m_1 \ll r_1 \ll b.
\ea
Equation~(\ref{frob}) is a generalized Frobenius series
\cite{Visser:2002ww}, where the expansion is about the regular
singular points $r_1=0$ and $r_1=\infty$.  There are clearly two
independent perturbation parameters, namely $\epsilon_{(3)}=m_1/r_1$
(the usual PN expansion parameter used in ${\cal{C}}_3$) and
$\epsilon_{(1)}=r_1/b$ (a tidal perturbation parameter used in
${\cal{C}_1}$). 
In the overlap region ${\cal{O}}_{13}$ we can expand in both.

%%%%%%%%%%%%%%%%%%%%%%%%%%%%%%%%%%%%%%%%%%%%%%%%%%%%%%
\section{Inner Zone Metric}
\label{inner}
%%%%%%%%%%%%%%%%%%%%%%%%%%%%%%%%%%%%%%%%%%%%%%%%%%%%%%

In this section we discuss the metric in the inner zone ${\cal{C}}_1$ of BH1
and its asymptotic expansion in the overlap region ${\cal{O}}_{13}$. 

Physically, we expect the spacetime in the inner zone of BH1 to be
Schwarzschild with mass $M_1$ plus a tidal perturbation due to BH2.
Thorne and Hartle~\cite{Thorne:1984mz} argue that, in the local
asymptotic rest frame (LARF) of BH1, the metric can be expanded in
powers of $M_1$ outside the horizon of BH1.  The first term,
independent of $M_1$, can be taken to represent the external universe
and thus can be computed by placing a test particle in the spacetime
of BH2 as done by Alvi~\cite{Alvi:1999cw}.  This is the tidal
perturbation due to BH2.  Terms of higher order in $M_1$ describe BH1
itself (the Schwarzschild metric) and interactions between BH1 and BH2
(tidally-induced quadrupole, etc).  At the level of approximation of
this paper, we can neglect the interaction terms because they are
$O(2,1)$ or higher.

\begin{figure}
\includegraphics[scale=0.5]{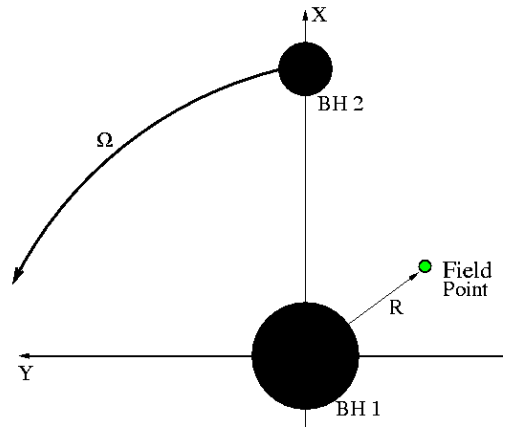}
\caption{\label{bhinner} Coordinate system used in inner zone 1. The
isotropic coordinates are centered on black hole $1$ and the other
hole orbits with angular velocity $\Omega$.  The matching
parameters $M_1$ and $\Omega$ and the coordinates $X^{\mu}$ need not
be equal to those used in the near zone.}
\end{figure}

Alvi identifies LARF coordinates, in terms of which the tidal
perturbation is obtained, with isotropic coordinates
(Fig.~\ref{bhinner}).
Observe that this coordinate system is centered on BH1 and is inertial.  
The asymptotic form of the inner-zone tidally perturbed metric valid in the
buffer zone is given by Eq.~(3.14) of Alvi~\cite{Alvi:1999cw}, who derives
it by extending Thorne-Hartle type arguments.
But to serve as initial data the perturbation is needed throughout all the
inner zone including the strong-field region, not just in the buffer zone
where the field is weak.
Alvi derives a form of the perturbation valid everywhere in the inner zone
and presents it in Eq.~(3.23) of Ref.~\cite{Alvi:1999cw}.
The result depends on parameters $M_1$ and $\Omega$ which will be related
to the near-zone parameters $m_1$ and $\omega$ when we perform the
matching.

We write Eq.~(3.23) of Ref.~\cite{Alvi:1999cw} as
\begin{widetext}
\ba
h_{00}^{IC,(1)} &\approx& -\left[{1-{M_1/(2 R_1)}}\over{1+{M_1/(2 R_1)}}\right]^2 
+ {m_2\over{b^3}} \left(1-{M_1\over{2 R_1}}\right)^4 
\left[ 3 \left( X \cos{\Omega T} + Y \sin{\Omega T} \right)^2 - R_1^2 \right],
\nonumber \\
h_{01}^{IC,(1)} &\approx& {2 m_2\over{b^3}} \sqrt{m\over{b}} \left(1-{M_1\over{2R_1}}\right)^2 
\left(1+{M_1\over{2R_1}}\right)^4 
\left[\left(Z^2-Y^2\right)\sin{\Omega T} - XY \cos{\Omega T}\right],
\nonumber \\
h_{02}^{IC,(1)} &\approx& {2 m_2\over{b^3}} \sqrt{m\over{b}} \left(1-{M_1\over{2R_1}}\right)^2 
\left(1+{M_1\over{2R_1}}\right)^4 
\left[\left(X^2-Z^2\right)\cos{\Omega T} + XY \sin{\Omega T}\right],
\nonumber \\
h_{03}^{IC,(1)} &\approx& {2 m_2\over{b^3}} \sqrt{m\over{b}} \left(1-{M_1\over{2R_1}}\right)^2  
\left(1+{M_1\over{2R_1}}\right)^4 
\left(Y \cos{\Omega T} - X \sin{\Omega T}\right) Z,
\nonumber \\
\label{internalmetricIC}
h_{ij}^{IC,(1)} &\approx& \left(1+{M_1\over{2 R_1}}\right)^4 \left( \delta_{ij} +
{m_2\over{b^3}} \left[3 \left(X \cos{\Omega T} + Y \sin{\Omega T}\right)^2 
- R_1^2 \right] \left\{ \left[ \left(1+{M_1\over{2 R_1}}\right)^4 
- {2 M_1^2\over{R_1^2}}\right] \delta_{ij} \right. \right. 
\nonumber \\
&& \left. \left. - {2 M_1\over{R_1}} \left(1 + {M_1^2\over{4 R_1^2}}\right)
{X^i X^j\over{R_1^2}} \right\} \right),
\ea
\end{widetext}
where once more the superscript $(1)$ is to remind us that this metric
is valid in submanifold ${\cal{C}}_1$, while the $IC$ superscript refers to
isotropic coordinates.
The uncontrolled remainders in all components of
Eq.~(\ref{internalmetricIC}) are $O(R_1/b)^3$.  

Inspection of Eq.~(\ref{internalmetricIC}) shows that the perturbed metric
diverges as $R_1\to0$ faster than $(1/R_1)^4$, which prevents the use of
puncture methods~\cite{Tichy:2002ec}.
Physically, this is because the tidal perturbation tacitly assumes a small
spacelike separation from the event horizon.
Of course this assumption is violated as $R_1\to\infty$, the asymptotically
flat spatial infinity we call the ``outside'' of hole 1; but it is also
violated as $R_1\to0$ since, in isotropic coordinates, that is the 
other asymptotically flat spatial infinity ``inside'' hole 1.
Outside the hole we match to the near zone metric which is well behaved,
but inside we have nothing to match to.
Thus numerical evolutions using our initial data will need to excise the
black holes.
Excision in practice still requires a slice that extends somewhat inside
each horizon, which raises the question of how far inside our data can be
considered valid.
Outside hole A the tidal perturbation is valid for $R_A \ll (m_A/m)b$.
The corresponding limit inside the hole is approximated by the
transformation $R_A \to (m_A/2)^2 /R_A$, which is an isometry for an
unperturbed Schwarzschild hole, where we have used $m_A \approx M_A$.
Thus the tidal perturbation is good roughly for
\begin{equation}
{m_A \over 2} {m \over 2b} \ll R_A \ll {m_A \over m} b,
\end{equation}
and the excision radius can be chosen anywhere between the lower limit of
this expression and the horizon.

Since matching will be simpler if performed between two coordinate
systems that live in charts that are similar to each other, we choose
to corotate first.  We define inner isotropic corotating coordinates
(ICC)
\ba
\bar{X} &=& X \cos{\Omega T} + Y \sin{\Omega T},
\nonumber \\
\bar{Y} &=& -X \sin{\Omega T} + Y \cos{\Omega T},
\nonumber \\
\bar{Z} &=& Z,
\nonumber \\
\bar{T}&=&T.
\ea
Using these equations, we can obtain the inner metric in isotropic
corotating coordinates, given by
\ba
\label{internalmetricICC}
h^{(1)}_{00} &\approx& H_t + H_{s1} \Omega^2 \left(\bar{X}^2+\bar{Y}^2\right) 
\nonumber \\
&& + 2 H_{st} \bar{X} {\Omega\over{b^2}} \left(\bar{X}^2+\bar{Y}^2-\bar{Z}^2\right),
\nonumber \\
h^{(1)}_{11} &\approx& H_{s1} - H_{s2} {\bar{X}^2\over{b^2}},
\nonumber \\ 
h^{(1)}_{01} &\approx&  -H_{s1}\bar{Y} \Omega - H_{st} {\bar{Y} \bar{X}\over{b^2}}, 
\nonumber \\ 
h^{(1)}_{22} &\approx& H_{s1}-H_{s2} {\bar{Y}^2\over{b^2}}, 
\nonumber \\ 
h^{(1)}_{02} &\approx& H_{s1} \bar{X} \Omega + {H_{st}\over{b^2}} \left(\bar{X}^2 
- \bar{Z}^2\right),
\nonumber \\
h^{(1)}_{33} &\approx& H_{s1}-H_{s2} {\bar{Z}^2\over{b^2}},
\nonumber \\ 
h^{(1)}_{03} &\approx& H_{st} {\bar{Y} \bar{Z}\over{b^2}},  
\nonumber \\ 
h^{(1)}_{12} &\approx& -H_{s2} {\bar{Y} \bar{X}\over{b^2}}, 
\nonumber \\
h^{(1)}_{23} &\approx& -H_{s2} {\bar{Z} \bar{Y}\over{b^2}},
\nonumber \\ 
h^{(1)}_{13} &\approx& -H_{s2} {\bar{Z} \bar{X}\over{b^2}},
\ea
where we use the shorthand              
\ba
H_{st} &=& 2 m_2 \sqrt{\frac{m}{b^3}} 
\left(1 - {M_1\over{2 R_1}}\right)^2 \left(1 + {M_1\over{2 R_1}} \right)^4,
\nonumber \\ 
H_{s1} &=& \left(1 + {M_1\over{2 R_1}}\right)^4 \left\{1 + 2{m_2\over{b^3}} 
R_1^2 P_2\left({\bar{X}\over{R_1}}\right)
\right. 
\nonumber \\
&& \left.
\left[\left(1 + {M_1\over{2 R_1}}\right)^4 - 2 {M_1^2\over{R_1^2}}\right]\right\}, 
\nonumber \\
H_{s2} &=& \left(1+{M_1\over{2R_1}}\right)^4 \left(1+{M_1^2\over{4R_1^2}}\right) {4m_2M_1\over{b R_1}} 
P_2\left({\bar{X}\over{R_1}}\right) ,
\nonumber \\ 
H_{t} &=& - \left({1-M_1/2 R_1}\over{1+M_1/2 R_1}\right)^2 + 2
\left(1-{M_1\over{2 R_1}}\right)^4
\nonumber \\
&& {m_2\over{b^3}} R_1^2 P_2\left({\bar{X}\over{R_1}}\right),
\ea
and the errors are still $O(R_1/b)^3$.  In
Eq.~(\ref{internalmetricICC}) we have dropped the superscript ICC in
favor of (1), which refers to submanifold ${\cal{C}}_1$.

By expanding Eq.~(\ref{internalmetricICC}) in powers of $M_1/R_1$,
which is permissible in overlap region $\mathcal{O}_{13}$, we obtain
\ba
\label{internalmetricAICC}
\tilde{h}_{00}^{(1)} &\sim&  -1 + {2 M_1\over{R_1}} + {2 m_2\over{b^3}} R_1^2 P_2
\left(\bar{X}\over{R_1}\right)
\nonumber \\
&& + \Omega^2 \left(\bar{X}^2+\bar{Y}^2\right), 
\nonumber \\
\tilde{h}_{01}^{(1)} &\sim& {-2 m_2\over{b^3}} \sqrt{\frac{m}{b}} \bar{Y} \bar{X} - \bar{Y} \Omega
\left(1 + {2 M_1\over{R_1}} \right),
\nonumber \\
\tilde{h}_{02}^{(1)} &\sim& {2 m_2\over{b^3}} \sqrt{\frac{m}{b}} \left(\bar{X}^2-\bar{Z}^2\right) 
+ \bar{X} \Omega \left(1 + {2 M_1\over{R_1}}\right),
\nonumber \\ 
\tilde{h}_{03}^{(1)} &\sim& {2 m_2\over{b^3}} \sqrt{\frac{m}{b}} \bar{Z} \bar{Y}, 
\nonumber \\ 
\tilde{h}_{ij}^{(1)} &\sim& \delta_{ij} \left[ 1 + {2 M_1\over{R_1}} + {2 m_2\over{b^3}}
R_1^2 P_2\left(\bar{X}\over{R_1}\right) \right]
\ea
where the errors are $O(2,3)$.  Like Eq.~(\ref{nearasympt}),
${\tilde{h}}_{\mu\nu}^{(1)}$ is a bivariate expansion in both
$\epsilon_{(1)}=R_1/b$ (valid in the inner zone ${\cal{C}}_1$) and
$\epsilon_{(3)} = M_1/R_1$ (valid in the buffer zone
${\cal{O}}_{13}$).  In other words, it is the asymptotic expansion in
the buffer zone to the asymptotic expansion in the inner zone.

%%%%%%%%%%%%%%%%%%%%%%%%%%%%%%%%%%%%%%%%%%%%%%%%%%
\section{Asymptotic Matching}
\label{matching}
%%%%%%%%%%%%%%%%%%%%%%%%%%%%%%%%%%%%%%%%%%%%%%%%%%

In this section, we concentrate on finding a matching condition
($\psi_{13}$) and a coordinate transformation ($\phi_{13}$) that maps
points in ${\cal{C}}_1$ labeled with isotropic corotating coordinates (ICC)
$\bar{X}^{\mu}$ to points in ${\cal{C}}_3$ labeled with harmonic corotating
coordinates (HCC) $x^{\mu}$. 
As already discussed, we concentrate on buffer zone ${\cal{O}}_{13}$,
while the matching condition $\psi_{23}$ and the coordinate
transformation $\phi_{23}$ in ${\cal{O}}_{23}$ will be given later by
a symmetry transformation. 

We assume that the coordinates are asymptotic to each other and that
they can be expanded in an implicit bivariate series.  That is, we assume
that the map $\phi_{13}: {\bar{X}}^{\mu} \to x^{\mu}$ has the form
\begin{widetext}
\ba
\bar{X} &\approx& \left(x-{m_2 b\over{m}}\right) 
+ x \left[ \left(m_2\over{b}\right)^{1/2} \chi_1(x^{\mu}) + 
\left(m_2\over{b}\right) \chi_2(x^{\mu})
+\left(m_2\over{b}\right)^{3/2} \chi_3(x^{\mu}) \right],
\nonumber \\
\bar{Y} &\approx& y \left[ 1 + \left(m_2\over{b}\right)^{1/2}  \gamma_1(x^{\mu}) 
+ \left(m_2\over{b}\right) \gamma_2(x^{\mu}) 
+ \left(m_2\over{b}\right)^{3/2} \gamma_3(x^{\mu}) \right],
\nonumber \\
\bar{Z} &\approx& z \left[ 1 + \left(m_2\over{b}\right)^{1/2} \zeta_1(x^{\mu}) 
+ \left(m_2\over{b}\right) \zeta_2(x^{\mu}) 
+ \left(m_2\over{b}\right)^{3/2} \zeta_3(x^{\mu}) \right],
\nonumber \\
\label{coordtransf}
\bar{T} &\approx& t \left[ 1 + \left(m_2\over{b}\right)^{1/2}  \tau_1(x^{\mu}) + 
\left(m_2\over{b}\right) \tau_2(x^{\mu}) 
+ \left(m_2\over{b}\right)^{3/2} \tau_3(x^{\mu}) \right] ,
\ea
\end{widetext}
where $\chi$, $\gamma$, $\zeta$ and $\tau$ are functions of the harmonic
corotating coordinates which do not depend on $m/b$ (or equivalently
$\omega$), but are power series in $r_1/b$.
We continue these power series only to $O(r_1/b)^2$ so that the errors here
in $\phi_{13}$ are $O(2,3)$ as in Eqs.~(\ref{nearasympt})
and~(\ref{internalmetricAICC}) which are linked by $\phi_{13}$.
The first term in the $\bar{X}$ equation above is chosen so that both
coordinate systems have their origins at the center of mass of the binary.

Like the coordinates, the matching parameters in the two coordinate systems
must be identical to lowest order in $m_2/b$.
(They must also be independent of coordinates.)
Then $\psi_{13}$ is given by 
\ba
\label{params}
M_1 &\approx& m_1 \left[1 + \left(m_2\over{b}\right)^{1/2} \eta_1 + {m_2\over{b}} \eta_2  
+ \left(m_2\over{b}\right)^{3/2} \eta_3 \right],
\nonumber \\
\Omega &\approx& \omega \left[1 + \left(m_2\over{b}\right)^{1/2} \kappa_1 
+ {m_2\over{b}} \kappa_2 + \left(m_2\over{b}\right)^{3/2} \kappa_3 \right],
\nonumber \\
\ea
where the errors are $O(m/b)^2$.

Before moving on with the calculation, let us discuss the physical meaning
of the assumptions we have just made.
Equation~(\ref{coordtransf}) implies that inner and near zone metrics are
identical in the buffer zone up to a change in coordinates. 
For a single black hole, in the buffer zone (which is outside the horizon),
the only difference between isotropic and harmonic coordinates is the
radial transformation
\begin{equation}
r = R \left( 1 + {M^2 \over 4R^2} \right),
\end{equation}
where $r$ is in harmonic coordinates and $R$ is in isotropic coordinates.
This has the asymptotic form posited in Eq.~(\ref{coordtransf}).
Thus the assumption of Eq.~(\ref{coordtransf}) is only needed in the buffer
zone for the matching in this section.
However, we will want to write our final results in a global coordinate
system which goes inside the horizons (though not all the way to the
asymptotically flat ends).
For this purpose we assume that the form of Eq.~(\ref{coordtransf}) holds
for all values of $r_1>0$.
This has the effect of defining a new coordinate system which is asymptotic
to harmonic coordinates in the near zone and to isotropic coordinates in
the inner zone.

Now let us return to the asymptotic matching. 
Using Eq.~(\ref{params}) we can transform Eq.~(\ref{internalmetricAICC}) to
harmonic corotating coordinates and impose the matching condition of
Eq.~(\ref{matchingcondition}),
\be
\label{ourmatch}
\tilde{g}_{\mu \nu}^{(3)}(x^{\gamma}) \sim 
\tilde{h}_{\alpha \beta}^{(1)}\left(\bar{X}^{\gamma}\left(x^{\gamma}\right)\right)
 {{\partial \bar{X}^{\alpha}}\over{\partial x^{\mu}}}
{{\partial \bar{X}^{\beta}}\over{\partial x^{\nu}}}.
\ee
Equation~(\ref{ourmatch}) provides $10$ independent asymptotic
relations per order, all of which must be satisfied simultaneously.
Each asymptotic relation results in a first-order partial differential
equation for the coordinate transformation, leading to $10$
integration constants per order.
As we shall see, these constants correspond to boosts, rotations, and
translations of the origin.

Equation (\ref{ourmatch}) must be solved iteratively in orders of
$(m/b)^{1/2}$. Evaluating the nonzero components (the diagonals) of
Eq.~(\ref{ourmatch}) at zeroth order in $m/b$, \textit{i.e.}\
comparing Eqs.~(\ref{nearasympt}) and~(\ref{internalmetricAICC}),
provides no information, since it only asserts that at lowest order
both metrics represent Minkowski spacetime.  This is true for any
matching formulation involving metrics of objects that would have
asymptotically flat spacetimes in isolation.

The asymptotic relations given by evaluating Eqs.~(\ref{nearasympt}),
(\ref{internalmetricAICC}), and~(\ref{ourmatch}) at $O(m/b)^{1/2}$ are
\ba
\chi_1 \sim -  x \chi_{1,x}, &\qquad& \zeta_1 \sim -z \zeta_{1,z}
\nonumber \\
\gamma_1 \sim - y \gamma_{1,y}, &\qquad& \tau_1 \sim - t \tau_{1,t},
\nonumber \\
t \tau_{1,x} \sim x \chi_{1,t}, &\qquad&  y \gamma_{1,t} -t \tau_{1,y} \sim \left(m_2\over{m}\right)^{1/2},
\nonumber \\
t \tau_{1,z} \sim z \zeta_{1,t}, &\qquad& x \chi_{1,y} \sim - y \gamma_{1,x}, 
\nonumber \\ 
\label{pdeD:1/2}
z \zeta_{1,x} \sim - x \chi_{1,z}, &\qquad& y \gamma_{1,z} \sim - z \zeta_{1,y},
\ea
where commas stand for partial differentiation.
The solution in terms of integration constants $C_i$ is
\ba
t\tau_1(x,y,z,t) &=& C_4x + C_5y - \sqrt{m_2/m}\,y + C_8z + C_9,
\nonumber \\
x\chi_1(x,y,z,t) &=& -C_1y + C_2z + C_4t + C_3,
\nonumber \\
y\gamma_1(x,y,z,t) &=& -C_1x + C_7z + C_5t + C_6,
\nonumber \\
z\zeta_1(x,y,z,t) &=& -C_2x - C_7y + C_8t + C_{10}.
\ea
For simplicity, we choose all $C_i=0$ except $C_5=\sqrt{m_2/m}$.
Thus the coordinate systems are identical at $O(m/b)^{1/2}$.
The coordinate transformation then becomes
\ba
\bar{X} &\approx& \left(x-{m_2 b\over{m}}\right) + x \left[ \left(m_2\over{b}\right) 
\chi_2(x^{\mu}) \right.
\nonumber \\ 
&& \left. +\left(m_2\over{b}\right)^{3/2} \chi_3(x^{\mu}) \right],
\nonumber \\
\bar{Y} &\approx& y \left[ 1 + \left(m_2\over{b}\right) \gamma_2(x^{\mu}) 
+ \left(m_2\over{b}\right)^{3/2} \gamma_3(x^{\mu}) \right],
\nonumber \\ 
&& + \sqrt{\frac{m_2}{b}} \sqrt{\frac{m_2}{m}} t,
\nonumber \\
\bar{Z} &\approx& z \left[ 1 + \left(m_2\over{b}\right) \zeta_2(x^{\mu})  
+ \left(m_2\over{b}\right)^{3/2} \zeta_3(x^{\mu}) \right], 
\nonumber \\ 
\bar{T} &\approx& t \left[ 1 + \left(m_2\over{b}\right) \tau_2(x^{\mu}) 
+ \left(m_2\over{b}\right)^{3/2} \tau_3(x^{\mu}) \right], 
\ea
where the errors are still $O(2,3)$. 

Applying asymptotic matching to $O(m/b)$, we obtain 
\ba
-\left(\tau_2 + t \tau_{2,t}\right) &\sim& 1-\frac{(x-m_2 b/m)}{b} = 1 - \frac{x_1}{b},
\nonumber \\
\chi_2 + x \chi_{2,x} &\sim& 1-\frac{(x-m_2 b/m)}{b} = 1 - \frac{x_1}{b},
\nonumber \\
\gamma_2 + y \gamma_{2,y} &\sim& 1-\frac{(x-m_2 b/m)}{b} = 1 - \frac{x_1}{b},
\nonumber \\
\zeta_2 + z \zeta_{2,z} &\sim& 1-\frac{(x-m_2 b/m)}{b} = 1 - \frac{x_1}{b},
\nonumber \\
x \chi_{2,t} - t \tau_{2,x} &\sim& {t\over{b}} + \left(m\over{m_2}\right)^{1/2} {y\over{b}} \kappa_1,
\nonumber \\
y\gamma_{2,t} - t \tau_{2,y} &\sim& -\left(m\over{m_2}\right)^{1/2} {x-m_2b/m\over{b}} \kappa_1,
\nonumber \\
\label{pdeD:1}
z \zeta_{2,t} &\sim& t \tau_{2,z},
\nonumber \\
x \chi_{2,y} &\sim& -y \gamma_{2,x},
\nonumber \\
x \chi_{2,z} &\sim& - z \zeta_{2,x},
\nonumber \\
\gamma_{2,z} &\sim& -z \zeta_{2,y}.
\ea
Once more we have a system of $10$ coupled partial differential
equations that now depends on $\kappa_1$, which is a parameter that
relates $\Omega$ and $\omega$.
For simplicity, we choose $\kappa_1=0$, so that the $\Omega=\omega$ to this
order.
The solution to this system is then given by
\begin{widetext}
\ba
\tau_2 &=& -\left[1-{x\over{b}}+\frac{m_2}{m}\right]
+ D_5 {y\over{t}} + D_4 {x\over{t}} + D_8 {z\over{t}} + {D_9\over{t}},
\nonumber \\
\chi_2 &=& 1 - {x\over{2b}} + \frac{m_2}{m} + {1\over{2xb}} \left(2 t^2
+ y^2 +z^2 \right) + D_1 {y\over{x}} + D_4 {t\over{x}} 
+ D_2 {z\over{x}} + {D_3\over{x}}, 
\nonumber \\
\gamma_2 &=& 1 - {x\over{b}} + \frac{m_2}{m} - D_1 {x\over{y}} + D_7
      {z\over{y}} + D_5 {t\over{y}} + {D_6\over{y}},
\nonumber \\
\zeta_2 &=& 1 - {x\over{b}} + \frac{m_2}{m} + D_8 {t\over{z}} -D_7 {y\over{z}} 
- D_2 {x\over{z}} + {D_{10}\over{z}},
\ea
where the $D_i$ are 10 more integration constants.
For simplicity we set them to zero, and the coordinate transformation becomes
\ba
\bar{X} &\approx& \left(x-{m_2 b\over{m}}\right) + x \left\{ \left(m_2\over{b}\right) 
\left[1 - {(x-2m_2 b/m)\over{2b}} \right] +\left(m_2\over{b}\right)^{3/2} \chi_3(x^{\mu}) \right\}
+ {m_2\over{2 b^2}} \left( 2 t^2 + y^2 +z^2 \right), 
\nonumber \\
\bar{Y} &\approx& y \left\{ 1 + \left(m_2\over{b}\right) \left[1 - {(x-m_2 b)/m\over{b}} 
\right]  + \left(m_2\over{b}\right)^{3/2} \gamma_3(x^{\mu}) \right\} + \sqrt{\frac{m_2}{b}} \sqrt{\frac{m_2}{m}} t,
\nonumber \\
\bar{Z} &\approx& z \left\{ 1 + \left(m_2\over{b}\right) \left[ 1-{(x-m_2 b)/m\over{b}}
 \right] + \left(m_2\over{b}\right)^{3/2} \zeta_3(x^{\mu}) \right\},
\nonumber \\
\label{match:1PN}
\bar{T} &\approx& t \left\{ 1 - \left(m_2\over{b}\right) \left[1-{(x-m_2 b/m)\over{b}}
 \right] + \left(m_2\over{b}\right)^{3/2} \tau_3(x^{\mu}) \right\},
\ea
with errors of $O(2,3)$. 

Now that matching has been completed to $O(m/b)^{0}$, $O(m/b)^{1/2}$
and $O(m/b)$, we can proceed with matching at $O(m/b)^{3/2}$.
However, keeping in mind our discussion in Sec.~\ref{scenario} of the
orders needed, we will only use the spatial-temporal part of the
asymptotic relations matrix~(\ref{ourmatch}),
\ba
x \chi_{3,t} -t \tau_{3,x} &\sim& -\left(m\over{m_2}\right)^{1/2} \left[{y\over{b}} 
\left(1- \kappa_2  - {4x\over{b}}  + 3 {m_2\over{m}} \right) \right],
\nonumber \\
t \tau_{3,y} - y \gamma_{3,t} &\sim& \left(m\over{m_2}\right)^{1/2} \Biggl( {x\over{b}} 
\left[\kappa_2 - 1 +  \frac{4 m_1 - 6 m_2}{m} - 4 \left(\frac{m_2}{m}\right)^2 + 8 \frac{\mu}{m} \right] 
+ \left(\frac{x}{b}\right)^2 \left( \frac{7}{2} - \frac{4 m_1}{m} +
  \frac{2 m_2}{m} \right) \Biggr.
\nonumber \\ 
&& \Biggl. + \left(\frac{y}{b}\right)^2 \left(-\frac{1}{2} -
  \frac{m_2}{m} + \frac{2 m_1}{m} \right)
+ \left(\frac{z}{b}\right)^2 \left(-\frac{3}{2} - \frac{m_2}{m} +
  \frac{2 m_1}{m} \right) + \left(\frac{t}{b}\right)^2 + \frac{4 \mu b}{m_2 r_1} 
\Biggr.
\nonumber \\
&& \Biggl. 
 + \left\{\frac{m_2}{m} \left[ 1 - \kappa_2 + 2 \left(\frac{m_2}{m}\right)^2 + 
\frac{3 m_2 - 4 \mu}{m} \right] - \frac{9\mu + 8m_1}{2m} + \frac{3}{2}
\right\} \Biggr), 
\nonumber \\
\label{pdeND:3/2}
t \tau_{3,z} - z \zeta_{3,t} &\sim&  \left(m\over{m_2}\right)^{1/2} {zy\over{b^2}}. 
\ea
\end{widetext}
For simplicity, we choose
$\kappa_2 = 1 + 3 m_2/m$. This completes the derivation of the
matching parameters, since $\eta_i$ and $\kappa_3$ did not appear in
the differential equations at all, and hence, we can neglect them to
this order. Note that this choice of parameter matching is different
from Alvi's choice, and thus our coordinate transformation is also
different.  Up to $O(m/b)^2$, the corresponding parameter matching
condition $\psi_{13}$ is
\be
\Omega \approx \omega \left[ 1 + {m_2\over{b}}\left(1 + 3 {m_2\over{m}} \right) \right], 
\quad M_1 \approx  m_1.
\ee
This choice of $\psi_{13}$ simplifies Eq.~(\ref{pdeND:3/2}), which now becomes
\begin{widetext}
\ba
t \tau_{3,x} - x \chi_{3,t} &\sim& -\left(m\over{m_2}\right)^{1/2} {4yx\over{b^2}}, 
\nonumber \\
t \tau_{3,z} - z \zeta_{3,t} &\sim&  \left(m\over{m_2}\right)^{1/2} {zy\over{b^2}}, 
\nonumber \\
t \tau_{3,y} - y \gamma_{3,t} &\sim& \left(m\over{m_2}\right)^{1/2} \Biggl( {x\over{b}} 
\left[ \frac{4 m_1 - 3 m_2}{m} - 4 \left(\frac{m_2}{m}\right)^2 + 8 \frac{\mu}{m} \right] 
+ \left(\frac{x}{b}\right)^2 \left( \frac{7}{2} - \frac{4 m_1}{m} +
  \frac{2 m_2}{m} \right) \Biggr.
\nonumber \\ 
&& \Biggl. + \left(\frac{y}{b}\right)^2 \left(-\frac{1}{2} -
  \frac{m_2}{m} + \frac{2 m_1}{m} \right)
+ \left(\frac{z}{b}\right)^2 \left(-\frac{3}{2} - \frac{m_2}{m} +
  \frac{2 m_1}{m} \right) + \left(\frac{t}{b}\right)^2 + \frac{4 \mu b}{m_2 r_1} 
\Biggr.
\nonumber \\
&& \Biggl. 
 + \left\{\frac{m_2}{m} \left[ 2 \left(\frac{m_2}{m}\right)^2 - 
\frac{4 \mu}{m} \right] - \frac{9\mu + 8m_1}{2m} + \frac{3}{2}
\right\} \Biggr).
\ea
As before, we choose the integration constants for simplicity (and to keep
the slicings close to each other), resulting in the following solution:
\ba
\tau_3(x,y,z,t) &=& 0,
\nonumber \\
\chi_3(x,y,z,t) &=& \left(m\over{m_2}\right)^{1/2}  {4 y t\over{b^2}},
\nonumber \\
\zeta_3(x,y,z,t) &=& -\left(m\over{m_2}\right)^{1/2}  { y t \over{b^2}}, 
\nonumber \\
\gamma_3(x,y,z,t) &=& -\left(m\over{m_2}\right)^{1/2} {t\over{y}}   \Biggl( {x\over{b}} 
\left[ \frac{4 m_1 - 3 m_2}{m} - 4 \left(\frac{m_2}{m}\right)^2 + 8 \frac{\mu}{m} \right] 
+ \left(\frac{x}{b}\right)^2 \left( \frac{7}{2} - \frac{4 m_1}{m} +
  \frac{2 m_2}{m} \right) \Biggr.
\nonumber \\ 
&& \Biggl. + \left(\frac{y}{b}\right)^2 \left(-\frac{1}{2} -
  \frac{m_2}{m} + \frac{2 m_1}{m} \right)
+ \left(\frac{z}{b}\right)^2 \left(-\frac{3}{2} - \frac{m_2}{m} +
  \frac{2 m_1}{m} \right) + \frac{1}{3} \left(\frac{t}{b}\right)^2 + \frac{4 \mu b}{m_2 r_1} 
\Biggr.
\nonumber \\
&& \Biggl. 
 + \left\{\frac{m_2}{m} \left[ 2 \left(\frac{m_2}{m}\right)^2 - 
\frac{4 \mu}{m} \right] - \frac{9\mu + 8m_1}{2m} + \frac{3}{2}
\right\} \Biggr).
\ea
\end{widetext}
To summarize, we have found a coordinate transformation $\phi_{13}$
and a set of parameter relations $\psi_{13}$ that produce asymptotic
matching to $O(3/2,3)$ in the $00$ and $ij$ components of the 4-metric
and to $O(2,3)$ in the $0i$ components.

Note however that the $\gamma_3$ piece of the coordinate
transformation becomes singular at $r_1=0$. 
Also recall that the point $r_1=0$ is not identical to 
the point $R_1=0$, where the inner zone metric perturbation
diverges. Hence if we excise the inner zone metric close to $R_1=0$,
the point $r_1=0$ might be outside the excised region, in which
case our coordinate transformation would introduce a coordinate
singularity outside the excised region. To get rid of this
singularity we will replace $r_1$ by
\be
\tilde{r}_{1} = \sqrt{r_1^2 + 6 m^2} .
\ee
This change amounts to adding a higher order term to the
coordinate transformation, which has no effect in the buffer zone at
the current order of approximation, but it has the advantage that the
resulting coordinate transformation is now regular at $r_1=0$. With
this replacement the coordinate transformation is given by
\begin{widetext}
\ba
\bar{X} &\approx& \left(x-{m_2 b\over{m}}\right) + x \left\{ \left(m_2\over{b}\right) 
\left[1 - {(x-2m_2 b/m)\over{2b}} \right] + \frac{m_2}{b} \sqrt{\frac{m}{b}} {4 y t\over{b^2}} \right\} 
+ {m_2\over{2 b^2}} \left( 2 t^2 + y^2 +z^2 \right),  
\nonumber \\
\bar{Y} &\approx& y \left\{ 1 + \left(m_2\over{b}\right) \left[1 - {x-m_2 b/m\over{b}} 
\right] \right\} - \frac{m_2}{b} \sqrt{\frac{m}{b}} t 
 \Biggl( {x\over{b}} 
\left[ \frac{4 m_1 - 3 m_2}{m} - 4 \left(\frac{m_2}{m}\right)^2 + 8 \frac{\mu}{m} \right] 
+ \left(\frac{x}{b}\right)^2 \left( \frac{7}{2} - \frac{4 m_1}{m} +
  \frac{2 m_2}{m} \right) \Biggr.
\nonumber \\ 
&& \Biggl. + \left(\frac{y}{b}\right)^2 \left(-\frac{1}{2} -
  \frac{m_2}{m} + \frac{2 m_1}{m} \right)
+ \left(\frac{z}{b}\right)^2 \left(-\frac{3}{2} - \frac{m_2}{m} +
  \frac{2 m_1}{m} \right) + \frac{1}{3} \left(\frac{t}{b}\right)^2 + \frac{4 \mu b}{m_2 \tilde{r}_{1}} 
 + \left\{\frac{m_2}{m} \left[ 2 \left(\frac{m_2}{m}\right)^2 - 
\frac{4 \mu}{m} \right] 
\right.
\Biggr.
\nonumber \\
&& \left. \Biggl. 
- \frac{9\mu + 8m_1}{2m} + \frac{3}{2} \right\} \Biggr)
+ \sqrt{\frac{m_2}{m}} \sqrt{\frac{m_2}{b}} t,
\nonumber \\
\bar{Z} &\approx&  z \left\{ 1 + \left(m_2\over{b}\right) \left[ 1-{x-m_2 b/m\over{b}}
 \right] - \frac{m_2}{b} \sqrt{\frac{m}{b}} {y t \over{b^2}} \right\},
\nonumber \\
\bar{T} &\approx& t \left\{ 1 - \left(m_2\over{b}\right) \left[1-{x-m_2 b/m\over{b}}
 \right] \right\},
\nonumber \\
\label{fulltransf}
\Omega &\approx& \omega \left[ 1 + {m_2\over{b}}\left(1+ 3 {m_2\over{m}} \right) \right], 
\nonumber \\
M_1 &\approx&  m_1.
\ea
\end{widetext}
The coordinate transformation for matching in the other overlap region
$\mathcal{O}_{23}$ is obtained by the following symmetry
transformation:  
\be
\label{symmetry}
1 \leftrightarrow 2,  \qquad x \to -x, \qquad    y \to -y, \qquad z \to z.
\ee

In Eq.~(\ref{fulltransf}), $t$ should be considered small just as $x$,
$y$, and $z$ are.  Recall that fundamentally the overlap regions are
4-volumes, although when we choose a time slicing we have to deal with
their projections on a spatial hypersurface ($m_A \ll r_A \ll b$).
Just as the overlap regions span a limited range of $r_A$, so they
span a limited range of $t$.  The post-Newtonian metric and the
perturbed Schwarzschild metric are formally stationary, but the true
physical system includes gravitational waves (not modeled here) which
for example change the orbital frequency on a radiation reaction
timescale.  While this timescale is longer than an orbital period,
which must be of order $b$, rotation and boosts mix space and time
terms and to be consistent with $r_A \ll b$ we must keep $t \ll b$.
Therefore we choose the
$t=0=T$ slice when discussing the approximate metric in the next
section, which restricts our overlap region to the intersection of
this 3-surface with the overlap 4-volume.

%%%%%%%%%%%%%%%%%%%%%%%%%%%%%%%%%%%%%%%%%%
\section{An approximate metric for binary black holes}
\label{secglobal}
%%%%%%%%%%%%%%%%%%%%%%%%%%%%%%%%%%%%%%%%%%

In this section we transform the inner zone metric in isotropic
corotating coordinates to harmonic corotating coordinates via
Eq.~(\ref{fulltransf}).
The metric in the inner zone of black hole $1$ is given by
\be
g_{\mu \nu}^{(1)} = h_{\bar{\delta} \bar{\rho}}^{(1)}
J^{\bar{\delta}}{}_{\mu} J^{\bar{\rho}}{}_{\nu}, 
\label{piecewisemetric}
\ee
where in the buffer zone the $g_{00}$ and $g_{ij}$ components have
errors of $O(2,3)$ and $g_{0i}$ has errors of $O(5/2,3)$. In the
above equation the Jacobian $J^{\bar{\mu}}{}_{\nu} = \partial_{\nu}
\bar{X}^{\bar{\mu}}$ can be expanded as
\begin{widetext}
\ba
\label{jacobian1}
J^{\bar{T}}{}_t &=& 1-{m_2\over{b}} \left(1-{x-m_2 b/m\over{b}} \right),
\nonumber \\ 
J^{\bar{T}}{}_x &=& {m_2 t \over{b^2}},
\nonumber \\
J^{\bar{X}}{}_t &=& {m_2\over{b}} \left({2t\over{b}} + 4{xy\omega\over{b}}\right),
\nonumber \\
J^{\bar{X}}{}_x &=& 1 + {m_2\over{b}} \left( 1 - {x-m_2 b/m\over{b}} + {4\omega y t\over{b}}\right),
\nonumber \\
J^{\bar{X}}{}_y &=& {m_2\over{b}} \left( {y\over{b}} + 4 {x \omega t\over{b}}\right),
\nonumber \\
J^{\bar{X}}{}_z &=& {m_2 z \over{b^2}},
\nonumber \\
J^{\bar{Y}}{}_t &=& \sqrt{m_2\over{m}} \sqrt{m_2\over{b}} - \frac{m_2}{b}
\sqrt{m\over{b}} \left\{ \frac{x}{b}
\left[\frac{4 m_1 - 3 m_2}{m} - 4 \left(\frac{m_2}{m}\right)^2 + 8 \frac{\mu}{m} \right]
+ \left(\frac{x}{b}\right)^2 \left( \frac{7}{2} - \frac{4 m_1}{m} +
  \frac{2 m_2}{m}\right) + \left(\frac{y}{b}\right)^2 \left(-\frac{1}{2} 
\right. \right.
\nonumber \\
&& \left. \left. 
- \frac{m_2}{m} +  \frac{2 m_1}{m} \right) \left(\frac{z}{b}\right)^2
\left(-\frac{3}{2} - \frac{m_2}{m} + \frac{2 m_1}{m} \right) +
\left(\frac{t}{b}\right)^2+ \frac{4 \mu b}{m_2 {\tilde{r}}_1} +
\frac{m_2}{m} \left[ 2 \left(\frac{m_2}{m}\right)^2 - 4 \frac{\mu}{m}
\right] - \frac{9 \mu + 8 m_1}{2 m} - \frac{3}{2} \right\},  
\nonumber \\
J^{\bar{Y}}{}_x &=& {-m_2\over{b}} {y\over{b}} - \frac{m_2}{b} \sqrt{m\over{b}}
\frac{t}{b} \left\{ \left[\frac{4 m_1 - 3 m_2}{m} - 4 \left(\frac{m_2}{m}\right)^2 + 8 \frac{\mu}{m} \right]
+ \frac{2 x}{b} \left( \frac{7}{2} - \frac{4 m_1}{m} + \frac{2
    m_2}{m} \right) - 4 \frac{\mu b^2}{m_2} \frac{x-m_2 b/m}{{\tilde{r}}_1^3} \right\},
\nonumber \\
J^{\bar{Y}}{}_y &=& 1 + {m_2\over{b}} \left(1-{x-m_2 b/m\over{b}}\right) -
\frac{m_2}{b} \sqrt{m\over{b}} t \left[
\frac{2y}{b^2} \left(-\frac{1}{2} - \frac{m_2}{m} + \frac{2 m_1}{m} \right) - 4 \frac{\mu b}{m_2}
\frac{y}{{\tilde{r}}_1^3} \right],
\nonumber \\
J^{\bar{Y}}{}_z &=& - \frac{m_2}{b} \sqrt{m\over{b}} t \left[\frac{2z}{b^2}
  \left(-\frac{3}{2} - \frac{m_2}{m} + 2 \frac{m_1}{m} \right) - 4
  \frac{\mu b}{m_2} \frac{z}{{\tilde{r}}_1^3}\right], 
\nonumber \\
J^{\bar{Z}}{}_t &=& -  \frac{m_2}{b} \sqrt{m\over{b}} {zy\over{b^2}},
\nonumber \\
J^{\bar{Z}}{}_x &=& -m_2 {z \over{b^2}},
\nonumber \\
J^{\bar{Z}}{}_y &=&  - \frac{m_2}{b} \sqrt{m\over{b}} {z t\over{b^2}},
\nonumber \\
J^{\bar{Z}}{}_z &=& 1 + {m_2\over{b}}\left(1-{x-m_2 b/m\over{b}}\right) -
\frac{m_2}{b} \sqrt{m\over{b}} {yt\over{b^2}}. 
\ea
\end{widetext}
Furthermore, $h_{\mu\nu}^{(1)}$ refers to the inner metric presented in
Eq.~(\ref{internalmetricICC}), where we substitute the coordinate
transformation given by Eq.~(\ref{fulltransf}) and
\be
R_1 =\left(\bar{X}^2 + \bar{Y}^2 + \bar{Z}^2\right)^{1/2}. 
\ee
The metric in the inner zone of black hole $2$ (${\cal{C}}_2$) is given by
the symmetry transformation~(\ref{symmetry}) applied to
Eq.~(\ref{piecewisemetric}).  

We now have all the ingredients to construct an approximate piecewise
metric, for example
\ba
\label{piece-metric}
g_{\mu \nu}^{piece} &\approx&  \left\{  \begin{array} {ll}
        g_{\mu \nu}^{(1)}, \qquad   r_1 < r_1^T
\\
        g_{\mu \nu}^{(2)}, \qquad   r_2 < r_2^T
\\
        g_{\mu \nu}^{(3)}, \qquad   r_1 > r_1^T, r_2 > r_2^T, r <
        \lambda/2 \pi
\end{array} \right.
\ea
for some transition radii $r_A^T$ which are chosen to be inside ${\cal
O}_{A3}$ for $A=\{1,2\}$.
In Eq.~(\ref{piece-metric}), $g_{\mu\nu}^{(3)}$ is given
in Eq.~(\ref{nearmetricC:1}), $g_{\mu \nu}^{(1)}$ is given in
Eq.~(\ref{piecewisemetric}) and $g_{\mu \nu}^{(2)}$ is the symmetry
transformed version of Eq.~(\ref{piecewisemetric}). The inner zone
pieces of this metric are accurate to $O(R_1/b)^2$, while the
post-Newtonian near zone pieces are accurate to $O(m/r)^{3/2}$.  In
the overlap region, the inner and near zone $0i$ components
asymptotically match up to $O(3/2,2)$ as required for the extrinsic
curvature, while all other components
asymptotically match only up to $O(1,2)$.

Note that when we applied the coordinate
transformation of Eq.~(\ref{fulltransf}) to the inner zone metric we kept
terms up to $O(m/b)^5$, which at first glance seems too high.
These terms are needed because the inner zone metric
represents a tidally perturbed black hole with errors in the physics
of $O[(m/b)(R_1/b)^3]$. 
Close to BH1 ($R_1 \sim m$), the error in the physics is
only of order $O(m/b)^4$ and hence we should keep terms to at least
order $O(m/b)^3$.  However, recall that the perturbed black hole
metric satisfies the Einstein equations up to errors of only 
$O[(m/b)^{5/2}(R_1/b)^2]$
even though its astrophysical resemblance to a binary black hole has
errors already at $O[(m/b)(R_1/b)^3]$. If we want to obtain a metric which is
close to the constraint hypersurface, we should keep terms larger than
$O[(m/b)(R_1/b)^3]$, but not larger than $O[(m/b)^{5/2}(R_1/b)^2]$.
In particular, close to the
black hole we have constraint violations of $O(m/b)^{9/2}$.  For example,
if we had dropped terms of order $O(m/b)^4$ in the inner zone metric
we would have introduced additional constraint violations at this
order.

%%%%%%%%%%%%%%%%%%%%%%%%%%%%%%%%%%%%%%%%%%%%%%%%%%%%%%%%%%%%%%%%%%%%
\subsection{Global character of the asymptotic metric}
%%%%%%%%%%%%%%%%%%%%%%%%%%%%%%%%%%%%%%%%%%%%%%%%%%%%%%%%%%%%%%%%%%%%

\begin{figure}
\includegraphics[angle=-90,scale=0.3]{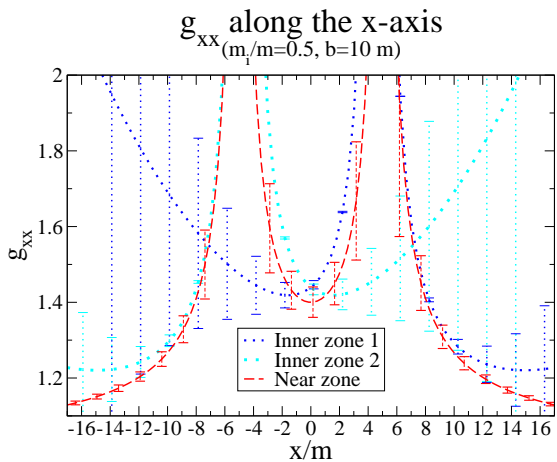}
\caption{\label{figure-metricXX}
This figure shows the $xx$ component
of the near zone metric (PN), denoted by a dashed line, and the
inner zone metrics (BHPT), denoted by dotted lines, along the
(harmonic) $x$ axis for a perturbative parameter $m/b=1/10$, with
the black holes located at $x/m = \pm 5$.
The buffer zones cannot be precisely defined, but most of the region
plotted is within one or the other (see text).
Matching does not guarantee that two curves which are asymptotically
matched intersect anywhere in the buffer zone, but rather that they are
comparable at the level of the uncontrolled remainders.
The error bars estimate these remainders as described in the text.
}
\end{figure}

\begin{figure}
\includegraphics[angle=-90,scale=0.3]{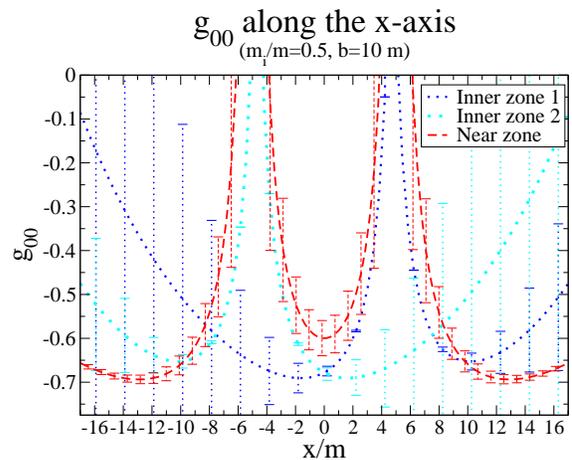}
\caption{\label{figure-metric00} This figure is similar to
Fig.~\ref{figure-metricXX}, but it shows the $00$ component of the
metric. Observe that in this component the differences between the
different approximations are more pronounced, although the general
features of asymptotic matching are still discernible.}
\end{figure}

In this subsection we plot Eq.~(\ref{piece-metric}) to describe some
features of the approximate piecewise metric. We choose a
system of equal-mass black holes $m_1=m_2=m/2$ separated by $b = 10
m$, so that both holes are located on the $x$ axis with BH1 at $x
\approx 5m$ and BH2 at $x\approx -5m$.
Figures~\ref{figure-metricXX}
and~\ref{figure-metric00} show the $xx$ and $00$ components of the
metric along the $x$ axis for this system.
(Other components of the piecewise $4$-metric exhibit similar behavior.)
In all plots we use a dashed line to denote the near
zone metric ($g_{\mu\nu}^{(3)}$) and dotted lines to denote the
inner zone metrics ($g_{\mu\nu}^{(1,2)}$)
We choose the separation $b=10m$ because it is near the minimum for which
our formalism makes sense, and we plot on the $x$-axis because it is where
the worst behavior occurs.
The idea is to (i) show some practical features of matching which have not
been presented in the literature and (ii) demonstrate the limits of the
method, particularly regarding the minimum separation.

In these plots, we also include error bars that estimate the
uncontrolled remainders in the approximations. These remainders can be
approximated by 
\begin{equation}
\label{PNerror}
2 [(m_1/r_1)^2 + (m_2/r_2)^2]
\end{equation}
in the near zone,
\begin{equation}
\label{BH1error}
2 (m_2/b) (R_1/b)^3
\end{equation}
in inner zone 1, and the same with $1\leftrightarrow2$ for inner zone 2.
The error in the near zone metric was estimated by looking at the next
order (2PN) term in the metric components~\cite{Blanchet:1998vx}.
[That term is much more complicated than Eq.~(\ref{PNerror}), but is
numerically about equal to it in the region plotted.]
The error in the inner zone metric comes about because the next tidal
correction for a single black hole of mass $m$ in perturbation theory will
be roughly proportional to ${1\over{3}} R_{i0j0,k}$. 
The error bars of the approximations are position-dependent, and thus
provide a useful sign of where each approximation is breaking down.
The PN error bars are larger near the holes than far away, and the BHPT
error bars exhibit the opposite behavior.
The error bars also provide an indicator of where both approximations are
comparably good:
Neglecting angular factors (as is typical in the literature), the error
bars for the near zone and inner zone $A$ are comparable at
\be
\label{Wolfs_rM}
r^T_A \approx \left( b^4 m_A^2 / m \right)^{1/5} ,
\ee
which takes a value of about $4.8m$ for the system
plotted here. This radius is a good candidate for the ``transition
radius'' of Eq.~(\ref{piece-metric}), but note that, in principle,
there is an infinite number of possible candidates, as long as they
are in the buffer zone. 
Furthermore, note that this is not a ``matching radius,'' since there is no
such thing.
Matching asymptotic expansions, as opposed to patching them as done by
Alvi~\cite{Alvi:1999cw}, does not happen at any particular place in the
buffer zone.
Rather, it makes two expansions comparable throughout the buffer zone up to
the uncontrolled remainders.

In Figs.~\ref{figure-metricXX} and~\ref{figure-metric00} we plot the $xx$
and $00$ metric components along the $x$-axis for the PN approximation as
well as for the two perturbed black hole approximations.
In Fig.~\ref{egg} the buffer zones around each black hole were sketched as
spherical shells around the holes, formally defined by $m_1 \ll r_1 \ll b$
and $m_2 \ll r_2 \ll b$.
It is important to recall that this definition, which is ubiquitous in the
literature, is imprecise because of the $\ll$ symbols and one cannot simply
substitute $<$ symbols.
(For one thing, there is angular dependence of the uncontrolled
remainders.)
Inserting the parameters of our system into these definitions, in
Fig.~\ref{figure-metricXX} the intersection of buffer zone 1 with the
$x$-axis is given by $5.5 \ll x/m \ll 15$ (to the right of BH1) and $-5 \ll
x/m \ll 4.5$ (to the left, between the holes).
In the part of the buffer zone to the right, away from BH2, we see a clean
example of the behavior expected of matched asymptotic expansions:
The near-zone curve and the BH1 inner-zone curve do not intersect, but the
difference between them is comparable to the estimated error bars
everywhere within this part of the buffer zone, even if we replace the
$\ll$ operator in the definition of the buffer zone by the precise $<$
operator.

Between the holes, to the left in Fig.~\ref{figure-metricXX}, the
interpretation of the curves is not so simple.
We cannot replace $\ll$ by $<$ in the definition of buffer zone 1 because
to the left of $x/m \approx 0$ ($r_1 \approx b/2$) the near-zone metric
component resembles that of BH2 rather than BH1.
This is because $m_1 \ll r_1 \ll b$ is a rough criterion obtained by
ignoring (among other things) the angular dependence of the expansion
coefficients in Eq.~(\ref{nearasympt}), inspection of which shows about a
factor of two variation as the angle is changed.
This angular dependence means that if one tries to redefine the buffer zone
heuristically as ``the region where the error bars on two curves overlap
and are not too large,'' it is significantly aspherical and can be squeezed
out entirely.
Even at the origin (where they are smallest), the error bars from the PN
approximation in the near zone are visibly larger than they are for most of
the right-hand part of buffer zone 1.
These error bars, however, are what is expected:
At the origin in Fig.~\ref{figure-metricXX} the $O(m/r)$ term which is kept
in the PN metric has a value of 0.4, for a total metric component of
$g_{xx}^{(3)} = 1.4$.
The uncontrolled $O(m/r)^2$ term we use for the error bar is 0.04, which is
precisely 10\% of the $m/b$ correction and about twice the value at
$x/m=\pm10$, a comparable distance on the other side of each hole.
The fact that the error bars are worst in between the holes does not depend
on $m/b$ or the mass ratio, but rather is a reflection of the physical
assumptions on which matching is based.
The near-zone metric is matched in buffer zone 1 to the metric of inner
zone 1, and in buffer zone 2 it is matched to the metric of inner zone 2,
but inner zone 1 is never matched to inner zone 2.
It is the intervening near zone that ensures that each black hole's tidal
perturbation is the appropriate one (up to uncontrolled remainders) for the
other black hole, because each tidal perturbation is derived for a black
hole without a nearby body.
(See, for example, the discussion in Sec.~II.B of Thorne and
Hartle~\cite{Thorne:1984mz}.)


\begin{figure}
\includegraphics[angle=-90,scale=0.3]{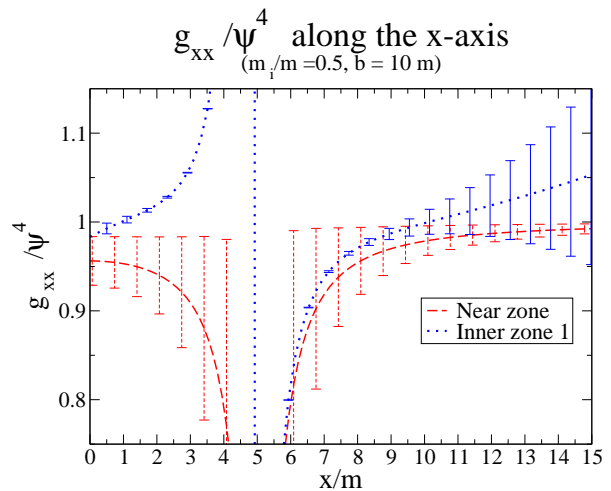}
\caption{\label{xx-Psi:4} In this figure we plot the $xx$ component of
  the near zone (PN - dashed line) and inner zone $1$ (BHPT -dotted line)
  metrics divided by the Brill-Lindquist factor
$\psi^4$. The behavior of the solutions
  is clearly different as we approach the event horizon from the left
(the direction of the other hole).
}
\end{figure}

Since the inner-zone and near-zone metric components diverge as $r_A\to 0$,
Fig.~\ref{figure-metricXX} might seem to imply that the solutions approach
each other near the horizons.
This misconception can be rectified by scaling the solutions to the
Brill-Lindquist factor $\psi^4$, where
\be
\psi = 1 + {m_1\over{2r_1}} + {m_2\over{2r_2}}.
\ee
This removes most of the divergent behavior of the solutions, as shown
in Fig.~\ref{xx-Psi:4}. In this figure, we only plot the region near
BH $1$ to show the difference in divergence better, but the region
near BH $2$ is very similar.
To the right, away from the other black hole, we see that the near-zone and
inner-zone solutions are indeed quite similar near BH1 and that there is a
wide region where both sets of error bars are comparable and overlap.
The transition radius $r^T_1 \approx 4.8m$ ($x/m \approx 10$) discussed
above is seen to be a good approximation of where the error bars are equal.
To the left, between the holes, there is only a small region (about $|x|/m <
1$) where the error bars overlap, and they are never equal.

The smallness of the region where the error bars overlap is an indication
that $b=10m$ is approaching the minimum separation for which our
approximation method makes sense.
The disappearance of such a region could be used as a criterion for the
failure of matching, although this is not a standard test and several
different approximate criteria could be used
(and this region is not the formal definition of a buffer zone anyway.)
In Fig.~\ref{figure-metric00} the error bars never do quite overlap at
$x/m \approx 0$, but they do overlap at $x/m \approx \pm 1$, although
we must keep in mind that they are rough estimates.
As discussed above, the error bars cannot be made equal at the origin by
changing $m/b$, although the overlap can be made better by increasing the
separation.
The fact that the rescaled metric components in Fig. \ref{xx-Psi:4} take off
in different directions as they approach BH1 from between the holes is
partly because the two metrics blow up at different coordinate locations.
This small relative translation is due to the coordinate system used in the
matching.

%%%%%%%%%%%%%%%%%%%%%%%%%%%%%%%%%%%%%%%%%%%%%%%%%%%%%%%%%%%%%%%%%%%%%%%%%%%%
\subsection{Transition Functions}
%%%%%%%%%%%%%%%%%%%%%%%%%%%%%%%%%%%%%%%%%%%%%%%%%%%%%%%%%%%%%%%%%%%%%%%%%%%%

The fact that matched curves do not strictly overlap even in the
buffer zone means that a piecewise metric such as 
Eq.~(\ref{piece-metric}) possesses discontinuities at the transition radii
$r^T_A$, wherever they are chosen to be.
These discontinuities can be problematic for numerical evolutions of the
spacetime and thus it is desirable to smooth them.
We now construct transition functions that smooth these discontinuities
out, by letting
\be
\label{quasiglobal}
g_{\mu \nu}^{(1, 3)} = \left[1-F_1(R_1)\right] g_{\mu\nu}^{(1)} 
                      + F_1(R_1) g _{\mu\nu}^{(3)},
\ee
where $F_1$ has the properties that 
$F_1(R_1 \gtrsim b)= 1$, 
$F_1(R_1 \lesssim m_1)= 0$, 
$F_1(R_1 \approx r_1^T)\approx 1/2$.
This ansatz yields a metric that is equal to the inner zone metric
near black hole 1, while it is equal to the near zone metric far
away from black hole 1. In between (i.e. in the buffer zone)
we obtain a weighted average of these two solutions. 
Since the Einstein equations are nonlinear, the sum of two solutions is in
general not another solution.
But since both solutions are valid in the buffer zone, and since
both have been matched, these two solutions are equal to each other
in the buffer zone up to uncontrolled remainders of $O(2,3)$,
corresponding to higher order post-Newtonian and tidal
perturbation terms. Hence any weighted average of these two solutions
in the buffer zone will yield the same correct solution up to uncontrolled
remainders of $O(2,3)$.
This justifies the use of smoothing to merge the two solutions in the 
buffer zone.
A similar solution can be obtained for the other black
hole by replacing $1 \to 2$.

We choose transition functions of the form
\ba
\label{AttBH}
f(r) = \left\{ \begin{array}{ll}
0 ,             \qquad r \leq r_0\\
\frac{1}{2} \left\{ 1 + \tanh \left[ \frac{s}{\pi} \left( \tan(\frac{\pi}{2w}(r-r_0)) 
\right. \right. \right. \\ 
\left. \left. \left. -\frac{q^2}{\tan(\frac{\pi}{2w}(r-r_0))}\right)
\right] \right\} ,      \qquad r_0 < r < r_0+w\\
1 ,\            \qquad  r \geq r_0+w .
\end{array} \right.
\ea
This function transitions from zero to one in a transition window which
starts at $r_0$ and has a width $w$.
The parameter $q$ controls the location at which $f$ reaches 1/2, and $s$
controls the slope at that location.
Note that this transition function is $C^{\infty}$, a property which is
useful for numerics.
In Eq.~(\ref{quasiglobal}) we set
\be F_A(R_A)=f(R_A), 
\ee
with parameters
\be 
r_0 = 1.5m, \qquad w = 5 r^T_A, \qquad q = 1/4, \qquad s = 10,
\ee
where $r^T_A$ is given by Eq.~(\ref{Wolfs_rM}).
With these parameters the transition functions reach the value $1/2$ at
$R_A \approx 1.5m+0.16w \approx 5.2m$, very close to $r^T_A \approx 4.8m$.
We could have used something more sophisticated such as a transition with
the same anisotropic behavior as the buffer zones, but several trials show
that such details do not matter as long as the transition function has the
right asymptotic properties.
(Trials also showed that the results were not too sensitive to the precise
parameter values.)
\begin{figure}
\includegraphics[angle=-90,scale=0.3]{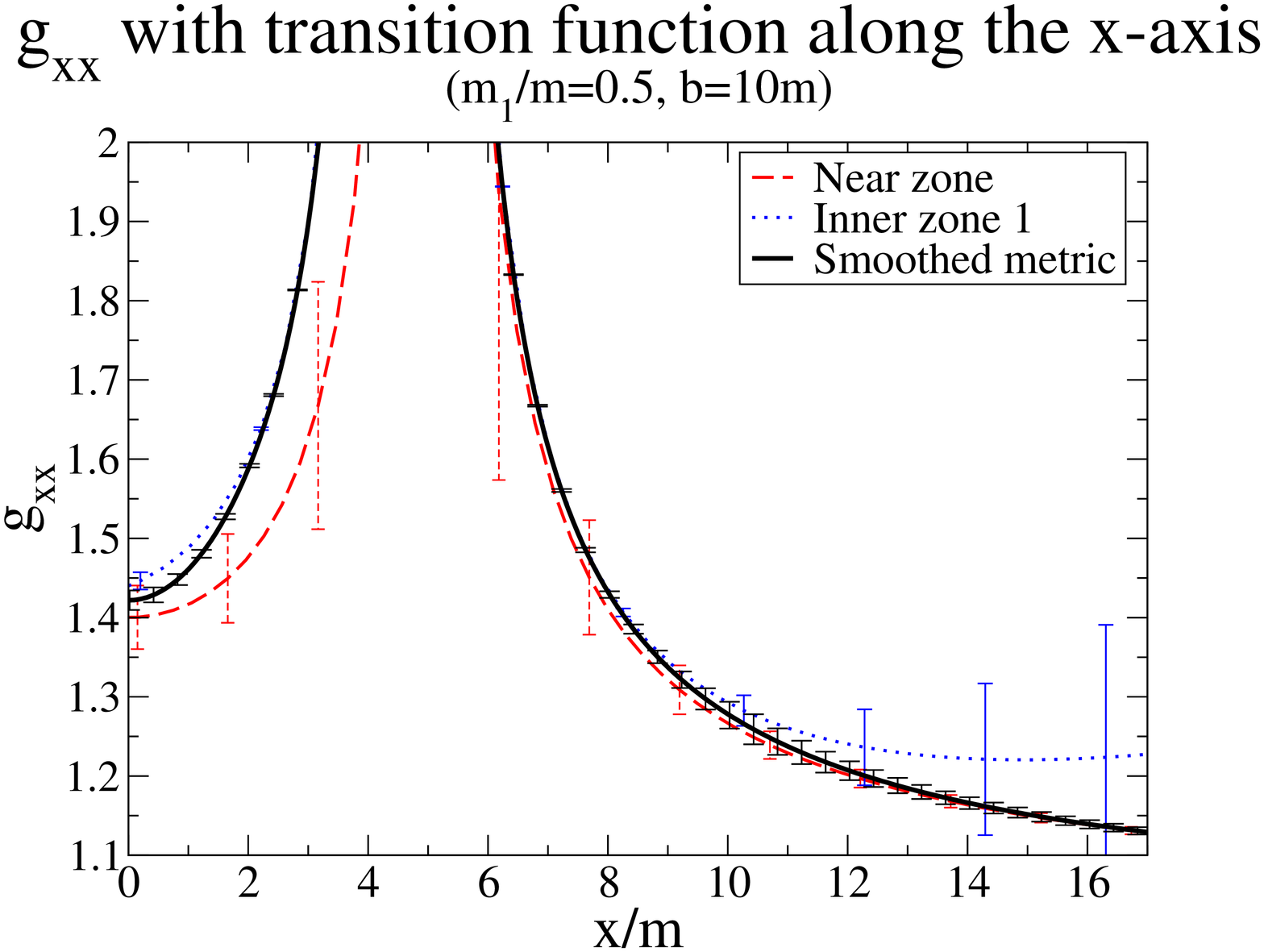}
\caption{\label{transitionfuncxx}
In this figure we show the $xx$ component of the global metric compared to
the near-zone PN metric and the inner-zone BHPT metric around BH1.
Observe that the transition function takes the global metric smoothly from
one to the other.
Error bars in the global metric are the same as the error bars of whichever
local approximation is better at that point.
}
\end{figure}

\begin{figure}
\includegraphics[angle=-90,scale=0.3]{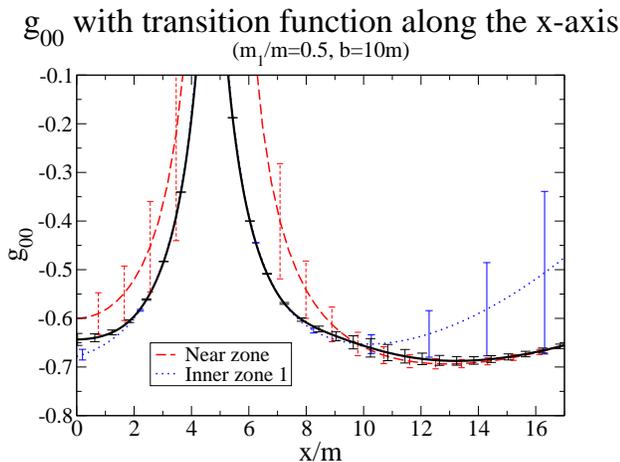}
\caption{\label{transitionfunc:00}
Same as the previous Figure, but for the $00$ metric component.
}
\end{figure}

We can see the effect of this transition function in
Fig.~\ref{transitionfuncxx} and~\ref{transitionfunc:00}. In these
figures we only show the region near BH $1$, but similar behavior is
observed near the other hole. The transition function effectively
takes one solution into the other smoothly in the buffer zone.
The size of the transition window can be
modified by changing the ``thickness'' $w$, but if $w$ is made too
small the derivatives of the metric (which include $1/w$ terms)
develop artificial peaks inside the transition window.

If the separation of the black holes is large enough so that the two
transition windows of width $w$ do not intersect, the transition
functions given in Eq.~(\ref{AttBH}) will suffice to generate a global
metric of the form
\ba
g_{\mu \nu}^{(global)} &=& F_2(R_2) F_1(R_1) g_{\mu \nu}^{(3)} +
\left[1-F_1(R_1)\right] g_{\mu \nu}^{(1)} 
\nonumber \\
&& + \left[1 - F_2(R_2)\right] g_{\mu \nu}^{(2)},
\ea
However, if the separation is small enough that the two transition windows
of width $w$ start to overlap, we must construct a third function $G(x)$
to allow for a smooth transition between the two black holes while not
contaminating the global solution near BH1 with a piece of the solution
from inner zone 2 and vice versa.
(In essence, this is an after the fact way of handling the fact that the
buffer zones are not really spherically symmetric as often implied in the
literature. When the holes are too close, a good transition function
should not be spherically symmetric either.) For the
system considered here, $b=10m$ is a sufficiently small separation that
such a third transition function is necessary. The global metric then
becomes
\ba
\label{global}
g_{\mu \nu}^{(global)} &=& G(x) \left\{F_1(R_1) g_{\mu \nu}^{(3)} +
\left[1-F_1(R_1)\right] g_{\mu \nu}^{(1)}\right\}
\nonumber \\
&& + \left[1-G(x)\right] \left\{F_2(R_2)g_{\mu \nu}^{(3)} + 
\left[1 \right.\right.
\nonumber \\
&& \left. \left. - F_2(R_2)\right] g_{\mu \nu}^{(2)}\right\}.
\ea
(Recall that $x$ is the distance along the axis between the holes with the
origin at the center of mass.)
The function $G(x)$ will be chosen such that it is equal to unity near
black hole $1$ and zero near black hole $2$. In between the two black
holes, $G(x)$ will range from zero to one, so that non-trivial
averaging will occur only in this region.  Again, this averaging is
allowed because the two solutions in the curly brackets are both
valid (and equal up to uncontrolled remainders) in between the two
black holes.  Near each black
hole we obtain the appropriate inner zone solution, while far away
$F_1(R_1)=F_2(R_2)=1$, so that we obtain the near zone solution
$g_{\mu \nu}^{(3)}$.  We choose the transition function $G(x)$ between
${\cal{C}}_1 \cup {\cal{C}}_3$ and ${\cal{C}}_2 \cup {\cal{C}}_3$ to
be of the same form as the function in Eq.~(\ref{AttBH}),
{\textit{i.e.}}
\be
G(x)=f(x) ,
\ee
but with different parameter values
\be
r_0 = \frac{b(m_2-m_1)}{2m} - \frac{b-m}{2} , \;
w   = b-m, \; 
q   = 1, \;
s   = 5/2 .
\ee
(The more complicated form of $r_0$ is to account for the origin of the $x$
coordinate being at the binary center of mass rather than on either hole.)

The reader might worry that the use of transition functions could introduce
large artificial gradients.
However, with a reasonable choice of transition functions this is not the case.
In order to understand why, let us look at the derivatives of the smoothed
metric in more detail.
Consider for example the overlap region $\mathcal{O}_{13}$ in which $F_2=1$
and the smoothed metric is thus $F_1 g_{\mu \nu}^{(3)} + \left[1-F_1\right]
g_{\mu \nu}^{(1)}$.
A derivative of this smoothed metric takes the form
\ba
\label{transy}
  F_1 g_{\mu \nu}^{(3)'} + \left[1-F_1\right] g_{\mu \nu}^{(1)'}
+ F_1' \left[g_{\mu \nu}^{(3)}-g_{\mu \nu}^{(1)}\right].
\ea
The last term is the worrying one since it involves a derivative of the
transition function.
But note that the coefficient multiplying this term is the difference
between the inner-zone and near-zone metrics in the buffer zone, and
therefore is by definition of the order of the uncontrolled remainders in the
expansions of the first two terms.
Therefore the third (unphysical) term will be safely absorbed into the
small uncontrolled remainders unless we make a pathological choice of
transition function---for example, one that has an inverse power of a small
expansion parameter built into it.
Our transition functions are chosen to avoid this.
Their maximum slope is roughly $1/r^T_A$, comparable to the slopes of
$g_{\mu \nu}^{(1)}$ and $g_{\mu \nu}^{(3)}$ in the vicinity of $r^T_A$
where the maximum occurs.
Thus the unphysical third term in Eq.~(\ref{transy}) is always formally
small.
(We demonstrate that it is also small in practice in Sec.~\ref{numerics}.)
The errors introduced into the constraint equations are found by
differentiating~(\ref{transy}).
Again, all terms involving derivatives of the transition functions are
multiplied by the differences between expanded metrics in the buffer zone,
which are of the same order as the uncontrolled remainders and therefore
can be neglected.

\begin{figure}
\includegraphics[angle=-90,scale=0.3]{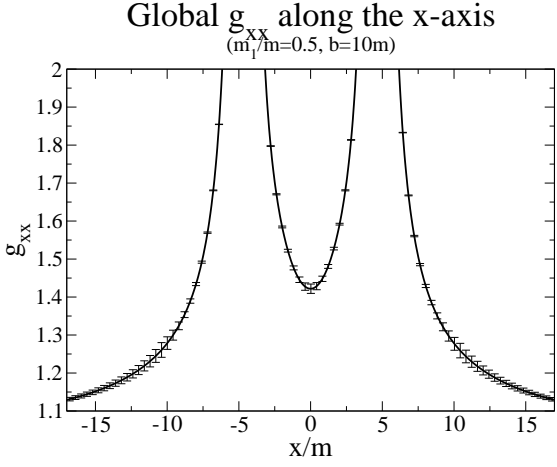}
\caption{\label{figure-smoothmetric} This figure shows the $xx$
component of the global approximation of the metric across both holes.
}
\end{figure}

\begin{figure}
\includegraphics[angle=-90,scale=0.3]{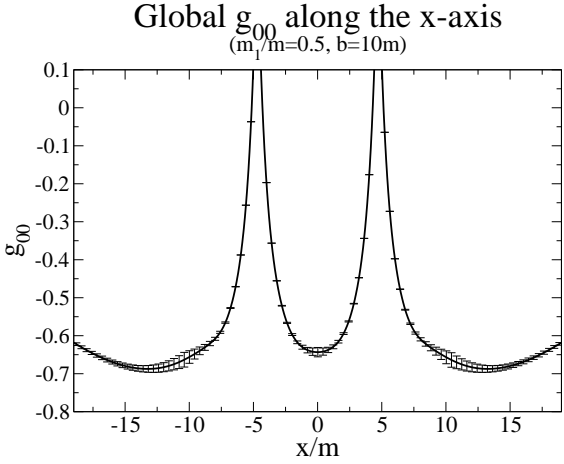}
\caption{\label{figure-smoothmetric:00} Same as the previous Figure, but
for the $00$ metric component.
}
\end{figure}

In Figs.~\ref{figure-smoothmetric} and~\ref{figure-smoothmetric:00}, we
demonstrate the good behavior of the smoothed solution.
Observe that as $x$ increases, the global metric with the transition
function becomes identical to the near zone metric, while as $r_A$
approaches zero (near each hole, $x/m\to\pm5$) it becomes equal to the
inner zone metric.
Since the holes are close to each other, in the region between the holes
the global metric never becomes identical to the near zone metric but
rather always contains a contribution from the inner zone metrics.
This linear combination is valid in that region because there the
asymptotic expansions of both approximate metrics are comparable to each
other. 
In Figs.~\ref{transitionfuncxx} and~\ref{transitionfunc:00}, the error bars
of the global metric with the transition function overlap the error bars of
the inner and near zone metrics in the regions where the former are valid.
This criterion would not be satisfied if the near zone and inner zone
curves were farther from each other, which would occur if the holes were
closer, and thus could be taken as another indicator that the holes are
still (barely) far enough apart for matching.

%%%%%%%%%%%%%%%%%%%%%%%%%%%%%%%%%%%%%%%%%%%
\section{Initial Data for Numerical Relativity} 
\label{numerics}
%%%%%%%%%%%%%%%%%%%%%%%%%%%%%%%%%%%%%%%%%%%
The approximate metric~(\ref{global}) could be used as initial data
for binary black hole simulations.  To facilitate this task we now
present this metric in the $3+1$ decomposition, by providing explicit
analytic expressions for the extrinsic curvature, lapse and shift on
the ${\bar{T}}=t=0$ slice. If the normal vector to this slice is
denoted by $n^a$, then the intrinsic metric in the slice is given by
\be q_{ab} = g_{ab} + n_a n_b, \ee and the extrinsic curvature is \be
K_{ab} = -\frac{1}{2} \pounds_n q_{ab} , \ee where $\pounds_n$ is the
Lie derivative in the direction normal to the $t=0$-slice.  Below we
compute $K_{ab}$ using the explicit expression \be K_{ab} =
-\frac{1}{2} n^c \left( \partial_c q_{ab} - \partial_a q_{cb} -
\partial_b q_{ac}\right), \ee which has been obtained using the
ordinary derivative operator and the fact that $q_{ab} n^b =0$.  The
evolution vector \be (\partial_t)^a = \alpha n^a + \beta^a \ee is
split into pieces perpendicular and parallel to the $t=0$ slice, where
$\alpha$ denotes the lapse and $\beta^a$ the shift. Note that $n^a n_a
=-1$ and $\beta^a n_a =0$.

The near zone extrinsic curvature computed from the PN metric
\cite{Blanchet:1998vx} in corotating harmonic coordinates on the $t=0$
slice is given by
\begin{equation}
K_{ij}^{(3)}=
\sum_{A=1}^{2} m_A \frac{ 4 v_{A}^{(i} n_{A}^{j)} 
                          - v_{A}^{k}  n^{A}_{k} \delta_{ij} }
                        { r_{A}^{2} },
\end{equation}
where the error is of $O(m/b)^{5/2}$, the superscript $3$ is
to remind us that this expression is only valid in ${\cal{C}}_3$, and
where the parentheses on the indices stand for symmetrization.  In
the previous equation, $v_{A}^{i}$ and $n_{A}^{i}$ denotes the
particle velocities and directional vectors given by
\begin{equation}
v_{1}^{2}=  \omega \frac{m_2}{m} b ,\qquad
v_{2}^{2}= -\omega \frac{m_1}{m} b ,\qquad
v_{A}^{1}=v_{A}^{3}=0
\end{equation}
and
\begin{equation}
n^{k}_{A} = \frac{x^k - x^k_{A}}{r_A} ,\ \ \
x^1_{1} =  \frac{m_2}{m} b , \ \
x^1_{2} = -\frac{m_1}{m} b ,\ \
x^{2}_{A} = x^{3}_{A}= 0  .
\end{equation}
The corresponding near zone lapse and shift on the $t=0$ slice are
\begin{equation}
\alpha_{(3)}= 1 - \sum_{A=1}^{2} \frac{ m_A }{ r_A }
\end{equation}
and
\begin{equation}
\beta_{(3)}^i= - \sum_{A=1}^{2} \frac{ 4 m_A v_A^i}{ r_A } 
               - \epsilon_{ij3} \omega x^j, 
\end{equation}
where once more this is valid on ${\cal{C}}_3$. 

The extrinsic curvature of the $\bar{T}=0$ slice valid in the inner
zone of black hole $1$ (${\cal{C}}_1$) and computed from the metric
given in the previous section in isotropic corotating coordinates is
\begin{widetext}
\ba
K_{00}^{(1),ICC} &=& -{m2\over{b^3}} \Omega^2 \bar{Y} \left(\bar{X}^2 + \bar{Y}^2\right) 
{\Psi^5\over{\Psi - \frac{M_1}{R_1}}} \left[2 \left(\Psi - \frac{M_1}{R_1}\right)^2 b \omega + 
3 \Omega \bar{X} \left(\Psi^4 - {2 M_1^2\over{R_1^2}}\right)\right],
\nonumber \\
K_{01}^{(1),ICC} &=& {m_2\over{b}} \Omega {\Psi^5\over{\Psi - \frac{M_1}{R_1}}} 
\left[3 \Omega \bar{X} {\bar{Y}^2\over{b^2}}
\left(\Psi^4 -{2M_1^2\over{R_1^2}}\right) + \left(\Psi - \frac{M_1}{R_1}\right)^2 
{\omega\over{b}} (\bar{X}^2 + 2 \bar{Y}^2) +
\left(\Psi - \frac{M_1}{R_1}\right) M_1 \omega {\bar{X}^2\over{R^3 b}} \right.
\nonumber \\
&& \left. \left(\bar{X}^2 + \bar{Y}^2 -  \bar{Z}^2 \right)\right],
\nonumber \\
K_{02}^{(1),ICC} &=& - {m_2 \bar{X}\bar{Y}\over{b^3}} \Omega {\Psi^5\over{\Psi - \frac{M_1}{R_1}}} 
\left[ \left(\Psi - \frac{M_1}{R_1}\right)^2 b \omega 
+ 3 \Omega \bar{X} \left(\Psi^4 - {2 M_1^2\over{R_1^2}}\right)- \left(\Psi - \frac{M_1}{R_1}\right)
 {b M_1\over{R_1^3}} \omega \left(\bar{X}^2 + \bar{Y}^2 - \bar{Z}^2\right)\right],
\nonumber \\
K_{03}^{(1),ICC} &=& {m_2\over{b^2}} \omega \Omega  \bar{X} \bar{Z} \Psi^5 \left[
{M_1\over{R_1^3}} \left(\bar{X}^2 + \bar{Y}^2 - \bar{Z}^2\right) -\Psi + \frac{M_1}{R_1} \right],
\nonumber \\
K_{11}^{(1),ICC} &=& - {m_2 \bar{Y}\over{2 b^3}} {\Psi^5\over{\Psi - \frac{M_1}{R_1}}} \left[ 
4 \left(\Psi - \frac{M_1}{R_1}\right)^2 b \omega + 4 \left(\Psi - \frac{M_1}{R_1}\right) 
b {M_1\over{R_1^3}} \omega \bar{X}^2 - 3 \Omega \bar{X} \left(4 {M_1^2\over{R_1^2}}
- 2 \Psi^4 + {M_1^3 \bar{X}^2\over{R_1^5}} \right.\right. 
\nonumber \\
&& \left. \left. 
+ {4 M_1 \bar{X}^2\over{R_1^3}}\right)\right],
\nonumber \\
K_{12}^{(1),ICC} &=& {m_2 \bar{X}\over{2 b^3}} {\Psi^5\over{\Psi - \frac{M_1}{R_1}}} 
\left[2 \left(\Psi - \frac{M_1}{R_1}\right)^2 b \omega + 12 M_1 \Omega 
{\bar{X} \bar{Y}^2\over{R_1^3}} \left(1 + {M_{1}^2\over{\left(\Psi - \frac{M_1}{R_1}\right) R_1^2}}
\right) + 2 \left(\Psi - \frac{M_1}{R_1}\right) b \omega {M_1\over{R_1^3}} \right. 
\nonumber \\
&& \left. 
\left(\bar{X}^2 - \bar{Y}^2 - \bar{Z}^2\right)\right],
\\ \nonumber
K_{13}^{(1),ICC} &=& {6 M_1 m_2 \Omega \bar{X}^2 \bar{Y} \bar{Z}\over{b^3 R_1^3}} 
{\Psi^5\over{\Psi - \frac{M_1}{R_1}}}\left(1 + {M_1^2\over{4 R_1^2}}\right),
\nonumber \\
K_{22}^{(1),ICC} &=& {m_2 \bar{Y}\over{2 b^3}} {\Psi^5\over{\Psi - \frac{M_1}{R_1}}} \left\{
12 {M_1^2\over{R_1^2}} \Omega \bar{X} - 6 \Omega \Psi^4 \bar{X} + 
3 {M_1^3 \bar{Y}^2\over{R_1^5}} \Omega \bar{X} + 4 {M_1\over{R_1^3}} \left[ 
3 \Omega \bar{X} \bar{Y}^2 + \left(\Psi - \frac{M_1}{R_1}\right) 
b \omega \left(\bar{X}^2 - \bar{Z}^2\right)\right]\right\},
\nonumber \\
K_{23}^{(1),ICC} &=& - {m_2 \bar{Z}\over{2 b^3}} {\Psi^5\over{\Psi - \frac{M_1}{R_1}}} 
\left[2 \left(\Psi - \frac{M_1}{R_1}\right)^2 b \omega - 
12 M_1 \Omega {\bar{X}\bar{Y}^2\over{R_1^3}} \left( 1 + {M_{1}^2\over{4 R_1^2}}\right) -
2 \left(\Psi - \frac{M_1}{R_1}\right) {b M_1 \omega\over{R_1^3}} \left(\bar{X}^2 
+ \bar{Y}^2 - \bar{Z}^2\right)\right],
\nonumber \\ \nonumber 
K_{33}^{(1),ICC} &=& {m_2 \bar{Y}\over{ 2 b^3}} {\Psi^5\over{\Psi - \frac{M_1}{R_1}}} \left[ 
4 \left(\Psi - \frac{M_1}{R_1}\right)^2 b \omega + 4 \left(\Psi - \frac{M_1}{R_1}\right) 
b \omega {M_1 \bar{Z}^2\over{R_1^3}} + 3 \Omega \bar{X} \left(4 {M_1^2\over{R_1^2}} 
- 2 \Psi^4 + {M_{1}^3 \bar{Z}^2\over{R_1^5}} + 4 {M_{1} \bar{Z}^2\over{R_1^3}}\right)\right],
\ea
\end{widetext}
where the error is of $O(R_1/b)^3$, and where the superscript $ICC$ is
to remind that that this is calculated in isotropic corotating
coordinates. Later on, we will transform this metric to harmonic
corotating coordinates with the map $\phi_{13}$ found in
Eq.~(\ref{fulltransf}), and we will drop this superscript.
In the above equations, $\Psi$ is the Brill-Lindquist factor for black
hole $1$ in isotropic coordinates, {\textit{i.e.}}
\be
\Psi = 1 + \frac{M_1}{2R_1}.
\ee

The $K_{0\mu}^{(1),ICC}$ components will be needed later in the
coordinate transformation and are obtained from the purely spatial
components $K_{kl}^{(1),ICC}$ using
\begin{equation}
K_{0\nu}^{(1),ICC} = q_{0}^{k} q_{\nu}^{l} K_{kl}^{(1),ICC},
\end{equation}
where the projection tensor $q_{\mu}^{\nu}$ is given by
\begin{equation}
q_{\mu}^{\nu} = \delta_{\mu}^{\nu} + n^{(1),ICC}_{\mu} n_{(1),ICC}^{\nu}.
\end{equation}
Here, the normal vector to the slice computed with $n^{(1),ICC}_{a}=
-\sqrt{-1/h^{00}_{(1)}} (d \bar{T})_{a}$ is given by
\begin{widetext}
\ba
n^{(1),ICC}_{0} &=& -\frac{{1-{M_1\over{2 R_1}}}}{\Psi} \left[ 1 - \frac{m_2}{b} 
\left(1-{M_1\over{2R_1}}\right)^2 \Psi^2 
\frac{R_1^2}{b^2} P_2({\bar{X} \over{R_1}}) \right] ,
\qquad  n^{(1),ICC}_{i} = 0 ,
\ea
where $P_2$ stands for the second Legendre polynomial.

The upper components are then 
\ba
n_{(1),ICC}^0 &=& \frac{\Psi}{1-{M_1\over{2 R_1}}} \left[ 1 + \frac{m_2}{b} 
\left(1-{M_1\over{2 R_1}}\right)^2 \Psi^2
\frac{R_1^2}{b^2} P_2(\bar{X} /R_1) \right] ,
\nonumber \\
n_{(1),ICC}^1 &=& 
\frac{\Psi}{1-{M_1\over{2R_1}}} \bar{Y} 
\left[ \Omega + 2\frac{m_2}{b} \omega \left(1-{M_1\over{2R_1}}\right)^2 
\frac{\bar{X}}{b}
+ \frac{m_2}{b} \Omega  \left(1-{M_1\over{2R_1}}\right)^2 
\Psi^2 \frac{R_1^2}{b^2} P_2(\bar{X} /R_1)             
\right],
\nonumber \\
n_{(1),ICC}^2 &=& -\frac{\Psi}{1-{M_1\over{2R_1}}} 
\left[ \Omega \bar{X} + 2\frac{m_2}{b} \omega \left(1-{M_1\over{2R_1}}\right)^2 
\frac{\bar{X}^2 - \bar{Z}^2}{b}
+ \frac{m_2}{b} \Omega  \left(1-{M_1\over{2R_1}}\right)^2 
\Psi^2 \frac{R_1^2}{b^2} \bar{X} P_2(\bar{X} /R_1) \right], 
\nonumber \\
n_{(1),ICC}^3 &=& -2\frac{m_2}{b} \omega \left(1-{M_1\over{2R_1}}\right)
 \Psi \frac{\bar{Y} \bar{Z}}{b} .
\ea
\end{widetext}
This means that the lapse and shift of the $\bar{T}=0$ slice in the
inner zone (${\cal{C}}_1$) and in inner corotating coordinates are
given by
\ba
\alpha_{(1),ICC} &=& -n^{(1),ICC}_0 
= \frac{1-{M_1\over{2R_1}}}{\Psi} \left[ 1 - \frac{m_2}{b} \left(1-
\right. \right.
\nonumber \\
&& \left. \left.  {M_1\over{2R_1}}\right)^2 \Psi^2 \frac{R_1^2}{b^2} P_2(\bar{X} /R_1) \right],
\nonumber \\
\label{alphaICC}
\beta_{(1),ICC}^i &=& -\alpha_{(1),ICC} n_{(1),ICC}^i.
\ea
Observe that the lapse of Eq.~(\ref{alphaICC}) goes through zero at
$R_1=M/2$.  Apart from a small perturbation it closely resembles the
standard Schwarzschild lapse in isotropic coordinates.

\begin{figure}
\includegraphics[angle=-90,scale=0.3]{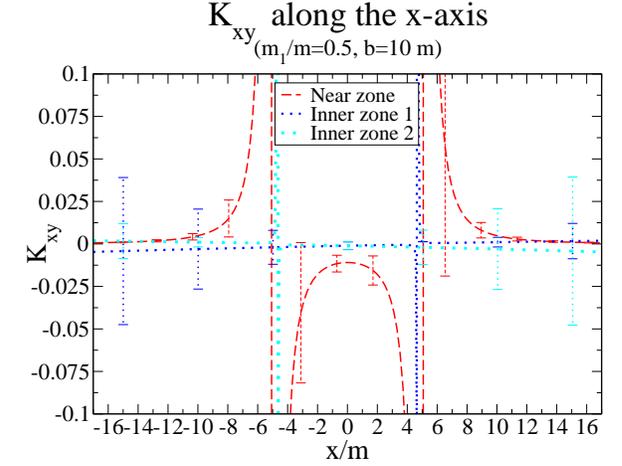}
\caption{\label{kxy} This figure shows the $xy$ component of the near
zone (PN - dashed line) extrinsic curvature, as well as the inner zone
curvatures (BHPT - dotted lines) obtained via black hole perturbation
theory. This figure uses the same test case as previous figures, with equal
mass black holes and $m/b=1/10$.}
\end{figure}

\begin{figure}
\includegraphics[angle=-90,scale=0.3]{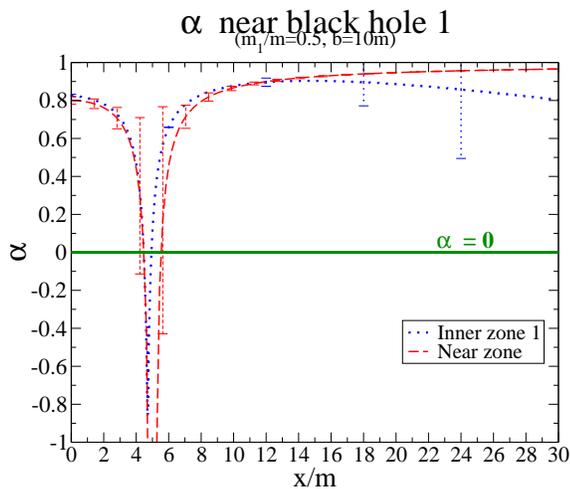}
\caption{\label{lapse_div} This figure shows the near zone (PN -
dashed line) and the inner zone lapse (BHPT - dotted line) along the
x-axis. Observe that the near zone lapse crosses zero at $x/m = 5.5$
and $x/m = 4.5$, which is the location of the event horizon in
harmonic coordinates.}
\end{figure}

\begin{figure}
\includegraphics[angle=-90,scale=0.3]{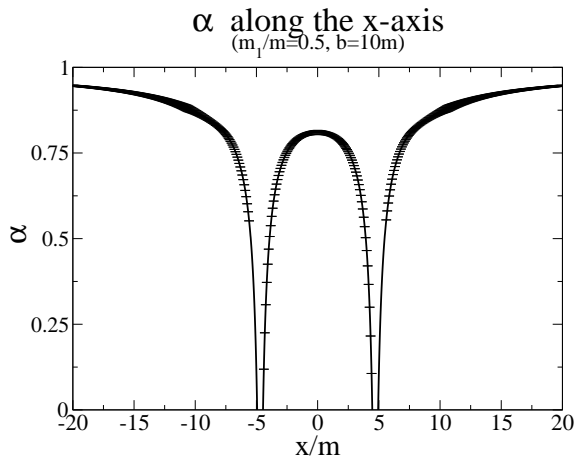}
\caption{\label{lapse_along_x} This figure shows the global lapse
along the x-axis with the transition function. }
\end{figure}

\begin{figure}
\includegraphics[angle=-90,scale=0.3]{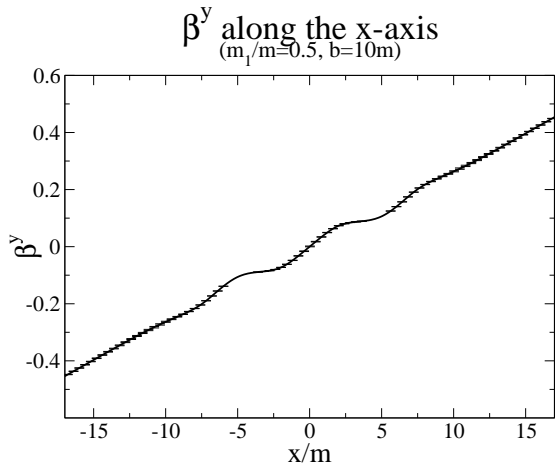}
\caption{\label{shift} This figure shows the global shift vector along
the x-axis with the transition function. }
\end{figure}

\begin{figure}
\includegraphics[angle=-90,scale=0.3]{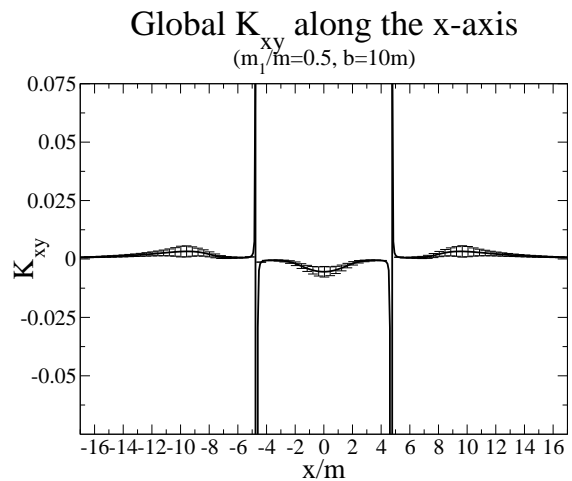}
\caption{\label{kxyglobal} This figure shows the global $xy$ component
of the extrinsic curvature along the x-axis with the transition
function.
Note that the bumps due to the transition function are comparable to the
error bars which estimate uncontrolled remainders in the expansions. }
\end{figure}

Note that the extrinsic curvature, lapse and shift given up to this point
are expressed in two different coordinate systems. 
The post-Newtonian quantities valid in the
near zone (${\cal{C}}_3$) are given in harmonic corotating
coordinates, while the black hole perturbation theory results valid in
the inner zones (${\cal{C}}_{1,2}$) are given in isotropic corotating
coordinates.  We will now apply the coordinate transformation found in
Sec.~\ref{matching}, namely Eq.~(\ref{fulltransf}), to transform the
inner zone expressions to harmonic corotating coordinates, thus
dropping the label ICC in favor of the superscript $(1)$.  The result
for the inner extrinsic curvature of black hole $1$ is given by
\be
K_{i j}^{(1)} = K_{\bar{l} \bar{m}}^{(1)}
J^{\bar{l}}{}_{i} J^{\bar{m}}{}_{j}, 
\ee
where all components still have errors of $O(5/2,3)$.  The extrinsic
curvature in harmonic corotating coordinates in submanifold
${\cal{C}}_2$ can be obtained from the above equation by the symmetry
transformation discussed in Eq.~(\ref{symmetry}).  In Fig.~\ref{kxy}
we have plotted the $xy$ component of the extrinsic curvature along
the $x$-axis. Observe that the behavior of the post-Newtonian solution
(dashed line) is different from that of the black hole perturbation
solution close to the black hole, where the latter diverges more
abruptly.
In Figs.~\ref{kxy}--\ref{kxyglobal}, the error bars have been estimated by
inserting Eqs.~(\ref{PNerror}) and~(\ref{BH1error}) into the definition of
the quantity plotted and using $\partial_t r_A \approx b \omega$ and
$\partial_t R_A \approx b \Omega$.

Similarly, the lapse and shift in harmonic corotating coordinates
corresponding to the inner zone of black hole $1$ are given by
\ba
\label{alphaHCC}
\alpha_{(1)}  &=& \frac{1}{ n_{(1)}^{0} } = - n^{(1)}_{0} =
J^{\bar{T}}{}_t \alpha_{(1),ICC},
\nonumber \\
\beta^i_{(1)} &=& -\alpha_{1} n_{1}^{i},
\ea
where again the lapse and shift for the inner zone around black hole
$2$ (${\cal{C}}_2$) can be obtained by the symmetry
transformation~(\ref{symmetry}).
In these equations, the normal vector is given by
\begin{equation}
n_{(1)}^{\mu} = J^{\bar{\nu}}{}_{\nu} n_{(1),ICC}^{\nu} ,
\end{equation}
where the matrix $J^{{\bar{\mu}}}{}_{\nu}$ has been defined in
Eq.~(\ref{jacobian1}).
Note that $\alpha_{(1)}$ in Eq.~(\ref{alphaHCC}) has the same zeros as
$\alpha_{(1),ICC}$ and, thus, $\alpha_{(1)}$ also changes sign at
$R_1=M/2$. Furthermore, since $J^{\bar{T}}{}_t = 1 + O(m/b)$ the inner zone
lapse $\alpha_{(1)}$ equals $\alpha_{(1),ICC}$ up to a perturbation of
$O(m/b)$ and thus $\alpha_{(1)}$ is equal to the standard lapse
of Schwarzschild in isotropic coordinates plus a perturbation of 
$O(m/b)$. These features are borne out by the plot in
Fig.~\ref{lapse_div} which shows the global lapse along the x-axis. We
can also see from the figure that while the inner zone lapse goes to
$-1$ as $r_1 \to 0$, the near zone one diverges to negative infinity.

With these equations, we can construct an approximate piecewise global
extrinsic curvature, lapse and shift by substituting the metric for
these quantities in Eq.~(\ref{piece-metric}). By the theorems of
asymptotic matching, the derivatives of adjacent pieces of the
piecewise metric will be asymptotic to each other inside their
respective buffer zones. This asymptotic similarity is, thus, also
observed in the extrinsic curvature, as well as the lapse and the
shift. Due to the piecewise nature of these solutions, there will be
discontinuities on a $2$-sphere located at some transition radius
inside of the buffer zone. In order to eliminate these
discontinuities, we use the same transition functions used for the
metric in Eq.~(\ref{AttBH}) with the same parameters. In this manner,
we obtain a smooth global extrinsic curvature, lapse and shift given
by
\begin{widetext}
\ba
K_{ij}^{(global)} &=& G(x) \left\{ F_1(R_1) K_{ij}^{(3)} + \left[1-F_1(R_1)\right] 
K_{ij}^{(1)}\right\} + \left[1-G(x)\right] \left\{ F_2(R_2) 
K_{ij}^{(3)}+  \left[1-F_2(R_2)\right] K_{ij}^{(2)}\right\},
\nonumber \\
\alpha_{(global)} &=& G(x) \left\{ F_1(R_1) \alpha_{(3)} + \left[ 1-F_1(R_1) \right] 
\alpha_{(1)}\right\} + \left[1-G(x)\right] \left\{ F_2(R_2) 
\alpha_{(3)}  +  \left[1-F_2(R_2)\right] \alpha_{(2)}\right\},
\nonumber \\
\label{globalextrinsic}
\beta^i_{(global)} &=& G(x) \left\{F_1(R_1) \beta^i_{(3)} + \left[1-F_1(R_1)\right] 
\beta^i_{(1)}\right\} + \left[1-G(x)\right] \left\{F_2(R_2)  
\beta^i_{(3)} +  \left[1-F_2(R_2)\right] \beta^i_{(2)}\right\}.
\ea
\end{widetext}
Figs.~\ref{lapse_along_x}, \ref{shift} and \ref{kxyglobal} show the
global lapse, shift and extrinsic curvature with the transition
functions.

We also could have computed the extrinsic curvature directly from
Eq.~(\ref{global}).  This would add terms involving derivatives of the
transition functions.  The parameters of the transition functions were
chosen so that these derivatives are of the same order as the
uncontrolled remainders in the buffer zone, and thus formally do not
affect the accuracy of the extrinsic curvature.  Since the two methods
are equivalent, we took the one which was simpler to compute (had fewer
terms).

The initial data constructed by the methods above [Eqs.~(\ref{global})
and~(\ref{globalextrinsic})] is only an approximate solution to the
Einstein equations. Therefore, this data leads to an error in the
constraints of the full theory of order $O(m/b)^{9/2}$ near the
horizons and $O(m/r)^2$ in the near zone. This error can be sufficiently
small compared to other sources of numerical error such that solving
the constraints more accurately is not required.  However, perhaps the
optimal approach would be to use this solution as input to York's
conformal method \cite{Cook:2000vr} and compute a numerical solution
to the full constraints. Since this data is already significantly
close to the constraint hypersurface, there might be some hope that
appropriate projection methods will {\textit{not}} alter much the
astrophysical content of the initial data. Somewhat surprisingly,
standard PN data (without matching) has not yet been used for the
generation of numerical black hole initial data except in
Ref.~\cite{Tichy:2002ec}, which is based on the PN data of
Ref.~\cite{Jaranowski:1997ky}. We leave it to future work to explore
similar techniques for the data set presented here.

If this data is to be evolved, it is necessary to choose a lapse and a
shift. The choice presented in this section is natural in the sense
that it is close to quasi-equilibrium. In other words, with the lapse
and shift presented in this section, the $3$-metric and extrinsic
curvature should evolve slowly. However, since our lapse is not
everywhere positive, some evolution codes may have trouble evolving
with it. If this is the case, one
can simply replace the above lapse with a positive function at the
cost of losing manifest quasi-equilibrium, but without changing the
physical content of the initial data or the results of the evolution.

%%%%%%%%%%%%%%%%%%%%%%%%%%%%%%%%%%%%%%%%%%
\section{Conclusions}
\label{conclusion}
%%%%%%%%%%%%%%%%%%%%%%%%%%%%%%%%%%%%%%%%%%

We have constructed initial data for binary black hole evolutions
by calculating a uniform
global approximation to the spacetime via asymptotic matching of
locally good approximations.  The manifold was first divided into
three submanifolds: two inner zones (one for each hole), ${\cal{C}}_1$
and ${\cal{C}}_2$ ($r_A \ll b$) equipped with isotropic coordinates;
and one near zone, ${\cal{C}}_3$ ($r_A \gg m_A$ and $r\ll \lambda/2
\pi$), equipped with harmonic coordinates.  In the near zone, the
metric was approximated with a post-Newtonian expansion, while in each
inner zone the metric was approximated with a perturbative tidal
expansion of Schwarzschild geometry.  Each approximate solution
depends on small parameters locally defined on each submanifold.
These submanifolds overlap in two buffer zones,
${\cal{O}}_{13}$ and ${\cal{O}}_{23}$ ($4-$volumes), given by the
intersection of the inner zones with the near zone, {\textit{i.e.}}
on an initial spatial hypersurface the buffer zones becomes
$3$-volumes given by $m_A \ll r_A \ll b$.  Inside each buffer zone,
two different approximations for the metric were simultaneously valid
and hence we were allowed to asymptotically match them inside this
zone.

The matching procedure consisted of first expanding both adjacent
approximate metrics asymptotically inside the buffer zones.
After transforming to the same gauge, these asymptotic
expansions were then set asymptotic to each other---equating their
expansion coefficients, which does not in general set the functions equal
to each other anywhere in the buffer zone.
After solving the differential
systems given by equating expansion coefficients,
asymptotic matching returned a coordinate transformation
($\phi_{13}$ and $\phi_{23}$) between submanifolds and matching
conditions ($\psi_{13}$ and $\psi_{23}$) that relate parameters native
to different charts.  A piecewise global metric was then obtained by
transforming all metrics with the set
$\left\{\phi_{nm};\psi_{nm}\right\}$, resulting in coordinates which
resemble harmonic coordinates in the near zone but isotropic coordinates in
each inner zone.

Once a piecewise global metric was found, the spatial metric and
extrinsic curvature were calculated in each zone by choosing a spatial
hypersurface, with the standard $3+1$ decomposition.  This initial
data was then transformed in the same manner as the $4$ metric.  Due
to the inherent piecewise nature of asymptotic matching, this data was
found to have discontinuities of order $O(3/2,3)$ or smaller inside
the buffer zone.  We constructed transition functions to remove the
remaining discontinuities in metric components and spikes in
derivatives. These transition functions were carefully built to avoid
introducing errors larger than the uncontrolled remainders of the
approximations in the buffer zones. With these functions, we
constructed a global uniform approximation to the metric valid
everywhere in the manifold with errors $O(m/b)^{9/2}$ near the black
holes and $O(m/r)^2$ far away from either of them. 

This uniform global approximation of the metric can be used as long as
the black holes are sufficiently far apart. When the two black holes
are too close, there is no
intervening post-Newtonian near zone in which to match. However, since
there is no precise knowledge of the region of convergence of the PN
series, it is unknown precisely at what separation the near zone
vanishes. We have experimented with separations $b \ge 10 m$ and we have
found that, in these cases, a region does exist between
the holes where the post-Newtonian metric is reasonably close to the
perturbed black hole metrics and thus matching is still possible.
For separations of $b
< 10 m$, this region shrinks rapidly and matching is
not guaranteed to be successful.
Also, the ``global'' metric is not valid all the way to the asymptotically
flat ends inside the holes, implying that our initial data must be evolved
with excision techniques rather than punctures.

We then constructed a lapse, shift, and extrinsic curvature,
all of which are needed for numerical evolutions.  The lapse was found
to possess the expected feature that it becomes negative inside the
horizon of either black hole.  Some numerical codes might find this
feature undesirable, in which case the lapse can be replaced by some
positive function at the cost of losing approximate
quasi-stationarity. These $3+1$ quantities were then smoothed
with transition functions of the same type as those used in the
$4$-metric.

In conclusion, we have constructed initial data for an inspiraling
black hole binary that satisfies the constraints to order
$O[(m/b)^{5/2}(R_1/b)^2]$ in the inner zone, and to order $O(m/r)^2$
in the near zone. This data is a concrete step
toward using PN and perturbation methods to construct such initial
data, and it should be compared to other numerical methods with
respect to its ability to approximate the astrophysical situation. We
should note that the data presented here makes use of perturbative
expansions of low order ({\textit{e.g.}}, the near zone metric is
built from a $1$ PN expansion), but this paper firms up a method
introduced by Alvi~\cite{Alvi:1999cw}
that could be repeated to higher order at the cost of more
algebra~\cite{Yunes:2006iw,higher-order-matching}. 
The post-Newtonian metric needed for the next order in $m/r$ (and beyond)
is available~\cite{Blanchet:1998vx}, as is the octopole perturbation
$(r/b)^3$ of a Schwarzschild black hole~\cite{Poisson:2005pi}.
Our method might also be extended to spinning
black holes, which are more astrophysically realistic, with the available
post-Newtonian near-zone metric~\cite{Tagoshi:2000zg} and tidal
perturbation~\cite{Yunes:2005ve}.

%%%%%%%%%%%%%%%%%%%%%%%%%%%%%%%%%%%%%%%%%%%%%%
\begin{acknowledgments}

We thank Thomas Baumgarte, Carl Bender, Lee Lindblom, Eric Poisson,
Kip Thorne, Qinghai Wang, and Clifford Will for useful discussions and
insightful comments.

We acknowledge the support of the Institute for Gravitational Physics
and Geometry and the Center for Gravitational Wave Physics, funded by
the National Science Foundation under Cooperative Agreement
PHY-01-14375. This work was also supported by NSF grants PHY-02-18750,
PHY-02-44788, PHY-02-45649, PHY-05-55628, PHY-05-55644, and by 
DFG grant ``SFB Transregio 7: Gravitational Wave Astronomy''.

\end{acknowledgments}
%%%%%%%%%%%%%%%%%%%%%%%%%%%%%%%%%%%%%%%%%%%%%%

%%%%%%%%%%%%%%%%%%%%%%%%%%%%%%%%%%%%%%%%%%%%%%%
% REFERENCES
%%%%%%%%%%%%%%%%%%%%%%%%%%%%%%%%%%%%%%%%%%%%%%%
\bibliography{paper.bib}

\end{document}